\documentclass[draftclsnofoot,onecolumn,12pt]{IEEEtran}
\usepackage{amsmath,amsfonts}
\usepackage{amssymb,amsbsy}
\usepackage[final]{graphicx}
\usepackage{times}
\usepackage{cite}
\usepackage{epsf,graphics,epsfig,subfigure}

\newcommand{\ignore}[1]{} 

\ignore{
\setlength{\topmargin}{-.5cm}
\setlength{\oddsidemargin}{-0.24cm}
\setlength{\evensidemargin}{-0.24cm}
\setlength{\headsep}{0pt}
\setlength{\headheight}{0pt}
\setlength{\textheight}{23cm}
\setlength{\textwidth}{17cm}
\setlength{\columnsep}{4mm}
}
\newtheorem{theorem}{Theorem}
\newtheorem{lemma}{Lemma}

\newtheorem{prop}{Proposition}

\newcommand{\mb}[1]{\boldsymbol{#1}} 

\newcommand{\hspp}{\hspace{0.02in}}

\newcommand{\indic}{\mbox{$1\!\!1$}}
\newcommand{\Sb}{\mb{S}}
\newcommand{\sbb}{\mb{s}}
\newcommand{\Zb}{\mb{Z}}
\newcommand{\qb}{\mb{q}} 
\newcommand{\pb}{\mb{p}}
\newcommand{\zb}{\mb{z}}

\begin{document}

\title{
Quickest Change Detection of a Markov Process Across a Sensor Array}
\author{{\hspace{0.15in}}Vasanthan Raghavan and Venugopal V.\ Veeravalli$^{\star}$
\thanks{The authors are with the Coordinated Science Laboratory and the 
Department of Electrical and Computer Engineering, University of 
Illinois at Urbana-Champaign, Urbana, IL 61801 USA. 
Email: {\tt \small \{vasanth,vvv\}@illinois.edu}. 
$^{\star}$Corresponding author.}
\thanks{This work has been supported in part by the U.S.\ Army Research 
Office MURI grant W911NF-06-1-0094, through a sub-contract from Brown 
University at the University of Illinois. This paper was presented in part 
at the 11th IEEE International Conference on Information Fusion, Cologne, Germany, 
2008~\cite{vasanth_fusion08}.} 
}


\maketitle

\markboth{SUBMITTED TO THE IEEE TRANSACTIONS ON INFORMATION 
THEORY,~DECEMBER 2008} {Raghavan {\emph{et al.}}:}

\baselineskip 18pt 

{\vspace{-0.3in}}
\begin{abstract} 
\noindent 
Recent attention in quickest change detection in the multi-sensor setting 
has been on the case where the densities of the observations change at the 
same instant at all the sensors due to the disruption. In this work, a more 
general scenario is considered where the change propagates across the sensors, 
and its propagation can be modeled as a Markov process. A centralized, 
Bayesian version of this problem, with a fusion center that has perfect 
information about the observations and {\em a priori} knowledge of the 
statistics of the change process, is considered. The problem of minimizing 
the average detection delay subject to false alarm constraints is formulated 
as a partially observable Markov decision process (POMDP). Insights into the 
structure of the optimal stopping rule are presented. 
In the limiting case of rare disruptions, we show that the structure of the 
optimal test reduces to thresholding the {\em a posteriori} probability of 
the hypothesis that no change has happened. We establish the asymptotic 
optimality (in the vanishing false alarm probability regime) of this 
threshold test under a certain condition on the Kullback-Leibler (K-L) 
divergence between the post- and the pre-change densities. In the special 
case of near-instantaneous change propagation across the sensors, this 
condition reduces to the mild condition that the K-L divergence be positive. 
Numerical studies show that this {\em low-complexity} threshold test results 
in a substantial improvement in performance over {\em naive} tests such as a 
single-sensor test or a test that wrongly assumes that the change propagates 
instantaneously. 
\end{abstract}


\section{Introduction} 
\label{s:intro} 
An important application area for distributed decision-making systems is 
in environment surveillance and monitoring. Specific applications 
include: i) Intrusion detection in computer networks and security 
systems~\cite{tart1,baras}, ii) monitoring cracks and damages to 
vital bridges and highway networks~\cite{agha}, iii) monitoring 
catastrophic faults to critical infrastructures such as water and gas 
pipelines, electricity connections, supply chains, etc.~\cite{cfp}, 
iv) biological problems characterized by an event-driven potential 
including monitoring human subjects for epileptic fits, seizures, 
dramatic changes in physiological behavior, etc.~\cite{farwell,ratnam}, 
v) dynamic spectrum access and allocation problems~\cite{li}, 
vi) chemical or biological warfare agent detection systems to protect 
against terrorist attacks, vii) detection of the onset of an epidemic, and 
viii) failure detection in manufacturing systems and large machines. 
In all of these applications, the sensors 
monitoring the environment take observations that undergo a change in 
statistical properties in response to a disruption (change) in the 
environment. The goal is to detect the point of disruption (change-point) 
as quickly as possible, subject to false alarm constraints.

In the standard formulation of the change detection problem, studied 
over the last fifty years, there is a sequence of observations whose 
density changes at some unknown point in time and the goal is to detect 
the change-point as soon as possible. Two classical approaches to quickest 
change 
detection are: i) The {\em minimax} approach~\cite{Lorden,Pollak}, where 
the goal is to minimize the worst-case delay subject to a lower bound on 
the mean time between false alarms, and ii) The {\em Bayesian} 
approach~\cite{Shiryaev1,Shiryaev2,Shiryaev3}, where the change-point 
is assumed to be a random variable with a density that is known 
{\em a priori} and the goal is to minimize the expected (average) 
detection delay subject to a bound on the probability of false alarm. 
Significant advances in both the minimax and the Bayesian theories of 
change detection have been made, and the reader is referred 
to~\cite{Shiryaev3,Shiryaev1,Siegmund,Tabook,BasNik,Lai_review,Taveer,poor_olympia,Shiryaev2,Lai,Page,Lorden,Pollak,Moustakides} 
for a representative sample of the body of work in this area. The 
reader is also referred 
to~\cite{BasNik,Beibel,Cdfuh,Lorden,Moustakides,Peskir,tart2,Taveer,yakir} 
for performance analyses of the standard change detection approaches in the 
minimax context, and~\cite{Lai_perf,TVAS} in the Bayesian context.

\ignore{ 
\begin{figure}[t]
\begin{center}
\setlength{\unitlength}{0.15in}
\begin{picture}(20,14)(2,-1)
\put(12,2){\oval(6.5,3)}
\put(12,0.5){\vector(0,-1){1.5}}
\put(12,-1.5){\makebox(0,0){change-point decision}}
\put(12,2.75){\makebox(0,0){Fusion}}
\put(12,1.25){\makebox(0,0){center}}
\put(4,8){\circle{2.8}}
\put(4,8){\makebox(0,0){$S_1$}}
\put(12,8){\makebox(0,0){Change}}
\put(8.5,8){\vector(1,0){1.5}}
\put(14,8){\vector(1,0){1.5}}
\put(12,8){\oval(4,2)}
\put(20,8){\circle{2.8}}
\put(20,8){\makebox(0,0){$S_L$}}
\put(4.87,6.86){\vector(4,-3){4.66}}
\put(19.11,6.86){\vector(-4,-3){4.66}}
\put(7,4.5){\makebox(0,0)[tr]{$Z_{k,1}$}}
\put(17,4.5){\makebox(0,0)[tl]{$Z_{k,L}$}}
\put(4, 11.25){\vector(0,-1){1.82}}
\put(20, 11.25){\vector(0,-1){1.82}}
\put(4,12){\makebox(0,0){$\{ Z_{k,1}\}$}}
\put(20,12){\makebox(0,0){$\{ Z_{k,L}\}$}}
\put(8,10.4){\circle*{.2}} \put(12,10.4){\circle*{.2}}
\put(16,10.4){\circle*{.2}}
\end{picture}
\end{center}
\caption{change-point detection across a sensor array \label{f:genst.fus}}
\end{figure}
}

\begin{figure}[htb!]
\centering
\begin{tabular}{c}
\includegraphics[height=3.2in,width=3.5in]{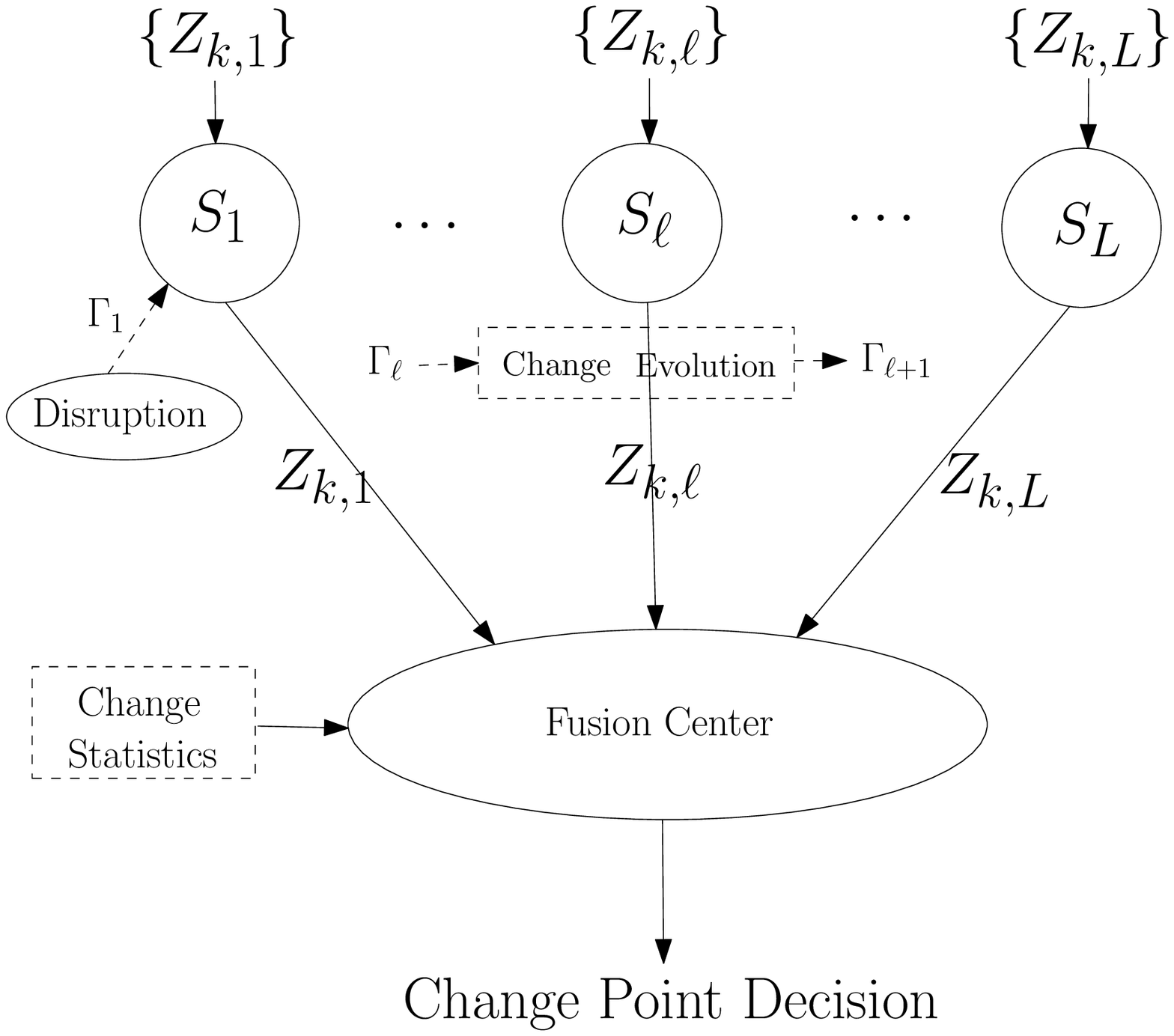}
\end{tabular}
\caption{Change-point detection across a linear array of sensors. 
\label{f:genst.fus}}
\end{figure}

Extensions of the above framework to the multi-sensor case where the 
information available for decision-making is {\em distributed} 
has also been explored~\cite{TVAS,TVfusion02,venu_fusion03,venu_qdecentral}. 
In this setting, the observations are taken at a set of $L$ distributed 
sensors, as shown in Fig.~\ref{f:genst.fus}. The sensors may send either 
quantized$/$unquantized versions of their observations or local decisions 
to a {\em fusion center}, subject to communication delay, power and 
bandwidth constraints, where a final decision is made, based on all the 
sensor messages. In particular, in recent 
work~\cite{TVAS,TVfusion02,venu_fusion03,venu_qdecentral}, it is assumed 
that the statistical properties of {\em all} the sensors' observations 
change at the same time. However, in many scenarios, it is more suitable 
to consider the case where the statistics of each sensor's observations may 
change at different points in time. An application of this model is in the 
detection of pollutants and biological warfare agents, where the change 
process is governed by the movement of the agent through the medium under 
consideration. Numerous other examples, including those described earlier, 
can be modeled in the change process detection framework. 

We consider a Bayesian version of this problem and assume that the point of 
disruption (that needs to be detected) is a random variable with a geometric 
distribution. We assume that the $L$ sensors are placed in an array or a line 
and they observe the change as it propagates through them. We model the 
inter-sensor delay with a Markov model and in particular, the focus is on 
the case where the inter-sensor delay is also geometric. More general 
inter-sensor delay models can be considered, but the case of a geometric 
prior has an intuitive and appealing interpretation due to the 
{\em memorylessness} property of the geometric random variable.

We study the centralized case, where the fusion center has complete 
information about the observations at all the $L$ sensors, the change 
process statistics, and the pre- and the post-change densities. This is 
applicable in scenarios where: i) the fusion center is geographically 
collocated with the sensors so that ample bandwidth is available for 
reliable communication between the sensors and the fusion center; 
and ii) the impact of the disruption-causing agent on the statistical 
dynamics of the change process and the statistical nature of the change 
so induced can be modeled accurately.

\noindent {\bf \em Summary of Main Contributions:} The goal of the 
fusion center is to come up with a strategy (or a stopping rule) to 
declare change, subject to false alarm constraints. Towards this goal, 
we first show that the problem fits the standard partially observable 
Markov decision process (POMDP) framework~\cite{bertsekas} with the 
sufficient statistics given by the {\em a posteriori} probabilities of 
the state of the system conditioned on the observation process. We then 
establish a recursion for the sufficient statistics, which generalizes 
the recursion established in~\cite{venu_qdecentral} for the case when 
all the sensors observe the change at the same instant. 

Following the logic of~\cite{venu_sequential} and~\cite{venu_qdecentral}, 
we then establish the optimality of a more general stopping rule for 
change detection. This rule takes the form of the smallest time of 
cross-over (intersection) of a linear functional (or hyperplane) in the 
space of sufficient statistics with a non-linear concave function, and 
generalizes the threshold test of~\cite{venu_qdecentral}. While further 
analytical characterization of the optimal stopping rule is difficult 
in general, in the extreme scenario of a rare disruption regime, we 
show that the structure of this rule reduces to a simple threshold test 
on the {\em a posteriori} probability that no change has happened. 
This low-complexity test is denoted as $\nu_{A }$ (corresponding to 
an appropriate choice of threshold $A$) for simplicity. 

While $\nu_{A }$ is obtained as a limiting form of the optimal test, 
it is not clear (as yet) if it is a ``good'' test. Towards this goal, 
we show that it is asymptotically optimal (as the false alarm 
probability $P_{\sf FA}$ vanishes) under a certain condition on the 
Kullback-Leibler (K-L) divergence between the post- and the pre-change 
densities. Meeting this condition becomes more easier as change 
propagates more instantaneously across the sensor array, and in the 
extreme case of~\cite{venu_qdecentral}, this condition reduces to the 
mild one that the K-L divergence be positive. 

The difference between the setting in this work and the setting 
in~\cite{venu_qdecentral} is in the non-asymptotic, but small $P_{\sf FA}$ 
regime. Asymptotic optimality 
of a particular test in the setting of~\cite{venu_qdecentral} 
translates to an $L$-fold increase in the slope of $E_{\sf DD}$ vs.\ 
$P_{\sf FA}$ in the regime where the false alarm probability is 
small, but not vanishing (e.g., $P_{\sf FA} \approx 10^{-4}$ or 
$10^{-5}$). However, if the change propagates too ``slowly'' across 
the sensor array, numerical studies indicate that 
not all of the $L$ sensors' observations {\em may} contribute to 
the performance of $\nu_{A }$ in this regime. Nevertheless, as 
$P_{\sf FA} \rightarrow 0$, all the $L$ sensors are expected (in general) 
to contribute to the slope. 

Thus, while it is not clear if $\nu_{A }$ is asymptotically optimal 
in general, or even if all the sensors' observations contribute to 
its performance in the non-asymptotic regime, numerical studies also 
show that it can result in substantial performance improvement 
over naive tests such as the {\em single sensor test} (where only the 
first sensor's observation is used in decision-making) or the 
{\em mismatched test} (where all the sensors' observations are used 
in decision-making, albeit with a wrong model that change propagates 
instantaneously), especially in regimes of practical importance 
(rare disruption, and reasonably quick, but non-instantaneous change 
propagation across the sensors). The performance improvement possible 
with $\nu_{A }$, in addition to its low-complexity, make it an attractive 
choice for many practical applications with a basis in multi-sensor 
change process detection.

\noindent {\bf \em Organization:} This paper is organized as follows. 
The change process detection problem is formally set-up in 
Section~\ref{sec2}. In Section~\ref{sec3}, this problem is posed in a 
POMDP framework and the sufficient statistics of the dynamic program 
(DP) are identified. Recursion for the sufficient statistics are then 
established. The structure of the optimal stopping rule in the general 
case and the rare disruption regime are illustrated in Section~\ref{sec4}. 
The limiting form of the optimal test is denoted as $\nu_{A}$ for 
simplicity. Using elementary tools from renewal theory, asymptotic 
optimality of $\nu_{A }$ is established in 
Sections~\ref{sectionx}--\ref{sectionxxx} under certain conditions. 
(The main results are stated in Sec.~\ref{sectionx} and they are established 
in detail in the appendices and in Sec.~\ref{sectionxx} and~\ref{sectionxxx}.) 
A discussion of the main results and numerical studies to illustrate our 
results are provided in Section~\ref{sec8}. Concluding remarks are made in 
Section~\ref{sec9}.

\section{Problem Formulation} 
\label{sec2}
Consider a distributed system with an array of $L$ sensors, as in 
Fig.~\ref{f:genst.fus}, that observes an $L$-dimensional discrete-time 
stochastic process $\Zb_k = \left[ Z_{k, 1}, \cdots, 
Z_{k, L} \right]$, where $Z_{k,\ell}$ is the observation at the $\ell$-th 
sensor and the $k$-th time instant. A disruption in the sensing environment 
occurs at the random time instant $\Gamma_1$ and hence, the 
density\footnote{We assume that the pre-change ($f_0$) and the post-change 
($f_1$) densities exist.} of the observations at each sensor undergoes a 
change from the null density $f_0$ to the alternate density $f_1$. 

\noindent {\bf \em Change Process Model:} Previous works on quickest 
change detection in multi-sensor systems consider strategies to detect the 
change-point, $\Gamma_1$, 
when the change occurs at the same instant across all the 
sensors~\cite{TVAS,TVfusion02,venu_fusion03,venu_qdecentral}. As 
described in the introduction, it is useful to consider more general 
scenarios where there exists random propagation delays in the change-point 
across the sensors. 

In this work, we consider a {\em change process} where the change-point 
evolves across the sensor array. In particular, the change-point as 
seen by the $\ell$-th sensor is denoted as $\Gamma_{\ell}$. We assume 
that the evolution of the change process is Markovian across the sensors. 
That is, 
\begin{eqnarray} 
P \big( \{ \Gamma_{\ell_1 + \ell_2 + \ell_3} = m_1 + m_2 + m_3 \} \big| 
\{ \Gamma_{\ell_1 + \ell_2} = m_1 + m_2 \}, 
\{ \Gamma_{\ell_1} = m_1 \} \big ) 
\nonumber \\ =  
P \big( \{ \Gamma_{\ell_1 + \ell_2 + \ell_3} = m_1 + m_2 + m_3 \} 
\big| \{ \Gamma_{\ell_1 + \ell_2} = m_1 + m_2 \} \big) \nonumber 
\end{eqnarray} 
for all $\ell_i$ and $m_i \geq 0, {\hspace{0.05in}} i = 1,2,3$. Further 
simplification of the analysis is possible under a {\em joint-geometric} 
model on $\{ \Gamma_{\ell} \}$. Under this model, the change-point 
($\Gamma_1$) evolves as a geometric random variable with parameter 
$\rho$, and inter-sensor 
change propagation is modeled as a geometric random variable 
with parameter $\{ \rho_{\ell - 1, \ell}, {\hspace{0.05in}} \ell = 
2, \cdots, L \}$. That is, 
\begin{eqnarray}
P \big(\{ \Gamma_1 = m \} \big) 
& = & \rho \left(1 - \rho \right)^{m}, 
{\hspace{0.05in}} m \geq 0 {\hspace{0.07in}} {\rm and} 
\nonumber \\
P \big( \{ \Gamma_{\ell} = m_1 + m_2 \} \big| 
\{ \Gamma_{\ell - 1} = m_2 \} \big) & = &  
\rho_{\ell - 1, \ell} {\hspace{0.03in}}  
( 1 - \rho_{\ell - 1, \ell} )^{m_1}, 
{\hspace{0.07in}}
m_1 \geq 0 \nonumber 
\end{eqnarray} 
independent of $m_2 \geq 0$ for all $\ell$ such that $2 \leq \ell \leq L$. 

We will find it convenient to set $\rho_{0,1} = \rho$ and $\rho_{L, L+1} = 0$ 
so that $\rho_{\ell-1, \ell}$ is defined for all 
$\ell = 1, \cdots, L+1$. 
This is also consistent with an equivalent $(L+2)$-sensor system 
where sensor indices run through $\{\ell = 0, \cdots, L+1 \}$. 
The hypothetical zero-th sensor 
models the disruption point, the first real sensor observes change 
with respect to the zero-th sensor with a geometric parameter $\rho$ 
(and so on). The hypothetical $(L+1)$-th sensor models an ``observer 
at infinity''\footnote{``Observer at infinity'' interpretations are 
often used in distributed decision-making and stochastic control 
problems~\cite{bertsekas,venu_sequential}.} 
that observes change from the $L$-th sensor with an infinite 
delay on average. This is reflected by setting $\rho_{L, L+1} = 0$. At 
this point, it should be noted 
that~\cite{TVAS,TVfusion02,venu_fusion03,venu_qdecentral} consider this 
equivalent framework {\em explicitly} by modeling $\gamma$, the probability that 
the disruption took place before the observations were made. The setup 
in~\cite{TVAS,TVfusion02,venu_fusion03,venu_qdecentral} can be obtained 
by setting: 
\begin{eqnarray}
P \big(\{ \Gamma_0 < 0  \}  \big) = \gamma {\hspace{0.07in}} {\rm and} 
{\hspace{0.07in}} P \big( \{\Gamma_0  = 0 \} \big) = 1 - \gamma 
{\hspace{0.07in}} {\rm for} {\hspace{0.07in}} {\rm some} 
{\hspace{0.07in}} \gamma \in [0,1]. 
\nonumber 
\end{eqnarray}
In this work, we focus on the case where $\gamma = 0$ with extension to 
the general case being straightforward.

%

While a joint-geometric model is consistent with the Markovian assumption 
as only the inter-sensory (one-step) propagation parameters are modeled, the 
change-points at the individual sensors themselves are {\em not} 
geometric. For example, it can be checked that 
\begin{eqnarray}
P\big( \{ \Gamma_2  = m \} \big) 
& = & \frac{ \rho {\hspace{0.03in}} \rho_{1,2} }
{ \rho - \rho_{1,2} } 
\times \Big( (1 - \rho_{1,2})^m - (1 - \rho)^m  \Big) 
\nonumber \\ 
P \big( \{ \Gamma_3 = m \} \big) & = & 
\frac{\rho {\hspace{0.03in}} 
\rho_{1,2} {\hspace{0.03in}} \rho_{2,3} } 
{ ( \rho - \rho_{1,2}  ) ( \rho_{1,2} - \rho_{2,3} ) ( \rho - \rho_{2,3} ) } 
\times \nonumber \\ 
& & {\hspace{-0.2in}} 
\Big( (\rho - \rho_{1,2} ) (1 - \rho_{2,3})^{m+2}  
- (  \rho - \rho_{2,3}) (1 - \rho_{1,2})^{m+2} + 
(\rho_{1,2} - \rho_{2,3} ) ( 1 - \rho )^{m+2}  \Big), 
\nonumber 
\end{eqnarray}
and so on. It should be clear from the above expressions that a 
joint-geometric model does not impose any constraints on 
$\{ \rho_{\ell-1, \ell} \}$ except that $\rho_{\ell-1,\ell} \in [0,1]$. 

Note that 
$\rho \rightarrow 1$ corresponds to the case where instantaneous 
disruption (that is, the event $\{ \Gamma_1 = 0\}$) has a high probability 
of occurrence. On the other hand, $\rho \rightarrow 0$ uniformizes 
the change-point in the sense that the disruption is equally likely to happen at 
any point in time. This case where the disruption is ``rare'' is of 
significant interest in practical 
systems~\cite{BasNik,poor_olympia,TVAS,TVfusion02,venu_fusion03,venu_qdecentral}. 
This is also the case where we will be able to make insightful statements about 
the structure of the optimal stopping rule. Similarly, we can also distinguish 
between two extreme scenarios at sensor $\ell$ depending on whether 
$\rho_{\ell-1,\ell} \rightarrow 0$ or $\rho_{\ell-1,\ell} \rightarrow 1$. The 
case where $\rho_{\ell-1,\ell} \rightarrow 1$ corresponds to instantaneous 
change propagation at sensor $\ell$ and $\{ \Gamma_{\ell} = \Gamma_{\ell -1} \}$ 
with high probability. The case where $\rho_{\ell-1,\ell} \rightarrow 0$ 
corresponds to uniformly likely propagation delay. The 
widely-used assumption~\cite{venu_qdecentral,TVAS} of instantaneous 
change propagation across sensors is equivalent to assuming 
$\rho_{\ell-1,\ell} = 1$ for all $\ell = 2, \cdots, L$.

\noindent {\bf \em Observation Model:} To simplify the study, we assume 
that the observations (at every sensor) are independent, 
conditioned\footnote{More general observation (correlation) models are 
important in practical settings. This will be the subject of future work.} 
on the change hypothesis corresponding to that sensor, and are identically 
distributed pre- and post-change, respectively. That is, 
\begin{eqnarray}
Z_{k, \ell } & \sim & \left \{ \begin{array}{ll}
{\rm i.i.d.} {\hspace{0.05in}} 
f_0 & {\rm if} {\hspace{0.1in}} k < \Gamma_{\ell}, \\ 
{\rm i.i.d.} {\hspace{0.05in}} 
f_1 & {\rm if} {\hspace{0.1in}} k \geq \Gamma_{\ell}.
\end{array} \right. 
\nonumber 
\end{eqnarray} 
We will describe the above assumption as that corresponding to an 
``i.i.d.\ observation process.'' Let $D(f_1, f_0)$ denote the 
Kullback-Leibler divergence between $f_1$ and $f_0$. That is, 
\begin{eqnarray} 
D(f_1, f_0) = \int \log \left( \frac{ f_1(x) }{ f_0(x) } \right) 
f_1(x) dx. 
\end{eqnarray}
We also assume that the measure described by $f_0$ is {\em absolutely 
continuous} with respect to that described by $f_1$. That is, if $f_1(x) = 0$ 
for some $x$, then $f_0(x) = 0$. This condition ensures that $E_{\bullet|f_1} 
\left[ \frac{ f_0(\bullet)} {f_1(\bullet) } \right] = 1$. 

\noindent {\bf \em Performance Metrics:} We consider a {\em centralized, 
Bayesian} setup where a fusion center has complete knowledge of the 
observations from all the sensors, $I_k \triangleq \{ \Zb_1, \cdots, 
\Zb_k \}$, in addition to knowledge of statistics of the change process 
(equivalently, $\{\rho_{\ell-1,\ell} \} $) and statistics\footnote{We assume 
that the fusion center has knowledge of $f_0$ and $f_1$ so that it can 
use this information to declare that a change has happened. Relaxing this 
assumption is important in the context of practical applications and is 
the subject of current work.} of the observation process (equivalently, 
$f_0$ and $f_1$). 
The fusion center decides whether a change has happened or not based on the 
information, $I_k$, available to it at time instant $k$ (equivalently, it 
provides a stopping rule or stopping time $\tau$). 

\ignore{ 
Let $P_k$ and $E_k$ denote the probability measure and the corresponding 
expectation when the change occurs at $\Gamma_1 = k$. Denote by 
$P_{\Gamma_1}$ the ``average'' probability measure defined as 
$P_{\Gamma_1}(\Omega) = P(\{ \Gamma_1 = m \}) P_k(\Omega)$, and 
$E_{\Gamma_1}$ denotes the expectation with respect to $P_{\Gamma_1}$. 
Two conflicting performance measures on change detection are the 
probability of false alarm, 
\begin{eqnarray} 
P_{\sf FA} \triangleq P_{\Gamma_1} \big( \{ \tau < \Gamma_1 \} \big) 
= \sum_{m=0}^{\infty} \rho (1 - \rho)^m P_m \big( \{ 
\tau < m  \} \big), \nonumber 
\end{eqnarray}
and the average detection delay, 
\begin{eqnarray}
E_{\sf DD} & \triangleq & E_{\Gamma_1} 
\left[ (\tau - \Gamma_1) \big| \tau \geq \Gamma_1 \right] 
= \frac{ E_{\Gamma_1} \left[ ( \tau - \Gamma_1  )^+  \right]  } 
{ P_{\Gamma_1} \big( \{ \tau \geq \Gamma_1 \}  \big)  }
\nonumber \\ 
& = & 
\frac{1}{  P_{\Gamma_1} \big( \{ \tau \geq \Gamma_1 \} \big) } 
\cdot 
\sum_{m = 0}^{\infty} \rho (1 - \rho)^m P_m \big( \{ \tau \geq m  \} 
\big) E_m\left[ (\tau - m ) \big| \tau \geq m  \right],
\nonumber 
\end{eqnarray} 
where $x^+ = \max(x,0)$. 
}

The two conflicting performance measures for quickest change detection 
are the probability of false alarm, $P_{\sf FA} \triangleq 
P \big( \{ \tau < \Gamma_1 \} \big)$, and the expected detection delay, 
$E_{\sf DD} \triangleq E \left[ (\tau - \Gamma_1)^+  \right]$, where 
$x^+ = \max(x,0)$. This conflict is captured by the Bayes risk, 
defined as, 
\begin{eqnarray}
R(c) & \triangleq & P_{\sf FA} + c E_{\sf DD} 
= E \left[ \indic \big( \{ \tau < \Gamma_1 \} \big) 
+ c \left ( \tau - \Gamma_1 \right)^+ \right] 
\nonumber 
\end{eqnarray} 
for an appropriate choice of per-unit delay cost $c$, where 
$\indic \big(\{ \cdot\} \big)$ is the indicator function of the event 
$\{ \cdot \}$. We will be particularly interested in the regime 
where $c \rightarrow 0$. That is, a regime where minimizing 
$P_{\sf FA}$ is more important than minimizing $E_{\sf DD}$, or 
equivalently, the asymptotics where $P_{\sf FA} \rightarrow 0$.

The goal of the fusion center is to determine 
\begin{eqnarray}
\tau_{\sf opt} = \arg\inf  \limits_{ \tau {\hspace{0.02in}} 
\in {\hspace{0.02in}} 
{\bf \Delta}_{\alpha} } E_{\sf DD}(\tau) \nonumber
\end{eqnarray} 
from the class of change-point detection procedures ${\bf \Delta}_{\alpha} 
= \big \{ \tau : P_{\sf FA}(\tau) \leq \alpha \big\}$ for which the 
probability of false alarm does not exceed $\alpha$. In other words, 
the fusion center needs to come up with a strategy (a stopping rule 
$\tau$) to minimize the Bayes risk. Note that the strategy developed by 
optimizing the Bayes risk can also be used for the other classical 
problem formulation in change detection, that of the minimax 
type~\cite[Theorem 1]{venu_qdecentral},~\cite{Shiryaev3,bertsekas}.

\section{Dynamic Programming Framework}
\label{sec3}
It is straightforward to check 
that~\cite[pp.\ 151-152]{Shiryaev3},~\cite{venu_qdecentral} the 
Bayes risk can be written as 
\begin{eqnarray}
R(c) & = & P \big( \{ \Gamma_1 > \tau \}  \big) + c 
E \left[ \sum_{k = 0}^{\tau - 1} 
P \big( \{  \Gamma_1 \leq k \} \big)  \right]. 
\nonumber 
\end{eqnarray} 
Towards solving for the optimal stopping time, we restrict attention to a 
finite-horizon, say the interval $[0, T]$, and proceed via a dynamic 
programming (DP) argument. 

The state of the system at time $k$ is the vector $\Sb_k  = 
[S_{k,1}, \ldots, S_{k,L}]$ with 
$S_{k,\ell}$ 
denoting the state at sensor $\ell$. The state $S_{k,\ell}$ can take 
the value 1 (post-change), 0 (pre-change), or $t$ (terminal). The system goes 
to the terminal state $t$, once a change-point decision $\tau$ has been declared. 
The state evolves as follows: 
\begin{eqnarray}
S_{k,\ell} & = & \indic \big( 
\{\Gamma_\ell  \leq k\}\cap \{S_{k-1,\ell} \neq t\} 
\cap \{\tau \neq k\} \big) 
+ t {\hspace{0.03in}} 
\indic \big( \{S_{k-1,\ell} = t\} \cup \{\tau = k \} \big) 
\nonumber 
\end{eqnarray}
with $\Sb_0 = {\bf 0}$. Since $\Sb_{k-1}$ captures the information contained 
in $\{ \Gamma_{\ell} \leq j \}$ for $0 \leq j \leq k-1$ and all $\ell$, 
given $\Sb_{k-1}$, $\{\Gamma_\ell \leq k\}$ is independent of 
$\{\Gamma_\ell \leq j, {\hspace{0.05in}} j \leq k-1\}$ 
for all $\ell$. Thus, the state evolution satisfies the Markov condition needed for 
dynamic programming. 

The state is not observable directly, but only through the observations. The 
observation equation can be written as
\[
Z_{k,\ell} = V_{k,\ell}^{(S_{k,\ell})} \indic 
\big( \{S_{k,\ell}\neq t\} \big) + 
\xi \indic \big( \{S_{k,\ell}= t\} \big), 
{\hspace{0.05in}} \ell \geq 1 
\]
where $V_{k,\ell}^{(0)}$ and $V_{k,\ell}^{(1)}$ are the $k$-th samples from 
independently generated infinite arrays of i.i.d.\ data according to $f_0$ and 
$f_1$, respectively. When the system is in the terminal state, the observations 
do not matter (since a change decision has already been made) and are hence 
denoted by a dummy random variable, $\xi$. It is clear that the 
observation uncertainty $(V_{k,\ell}^{(0)}, V_{k,\ell}^{(1)})$ satisfies the 
necessary Markov conditions for dynamic programming since they are i.i.d.\ in time.

Finally, the expected cost (Bayes risk) can be expressed as the expectation of 
an additive cost over time by defining
\[
g_k (\Sb_k) = c \indic \big( \{S_{k,1}  = 1\} \big) 
\]
and a terminal cost $\indic \big( \{S_{k,1} = 0 \} \big)$. Thus the 
problem fits the standard POMDP framework with termination~\cite{bertsekas}, 
with the sufficient statistic (belief state) being given by 
\[
P 
\big( \{\Sb_k = \sbb_k \} | I_k \big), 
\] 
where $I_k = \{ \Zb_1, \ldots, \Zb_k \}$ for $k$ such that $\Sb_k \neq t$, i.e., 
$S_{k,\ell} \in \{0,1\}$ for each $\ell$.  Note that this sufficient statistic 
is described by $2^{L}$ conditional probabilities, corresponding to the $2^{L}$ 
values that $\sbb_k$ can take.  We will next see that this sufficient statistic 
can be further reduced\footnote{This should not be entirely surprising 
since there exists a ``natural'' ordering on the sensors' change-points. 
They can be arranged in non-decreasing order: $\Gamma_{\ell} \geq 
\Gamma_{\ell-1}$ for all $\ell$. The primary reason for such an ordering 
to exist is that we assume an array (or line) of sensors in this work. 
Extensions to more general (or unknown) geometries of sensors is of 
interest in practice.} to only $L$ 
independent probability parameters in the general case. 

The fusion center determines $\tau$ and hence, the minimum expected 
cost-to-go at time $k$ for the above DP problem can be seen to be a 
function of $I_k$. For a finite horizon $T$, the cost-to-go function is 
denoted as $\widetilde{J}_k^T(I_k)$ and is of 
the form (see~\cite{venu_qdecentral},~\cite[p.\ 133]{bertsekas} for 
examples of similar nature): 
\begin{eqnarray} 
\widetilde{J}_T^T(I_T) 
& = & P \big( \{ \Gamma_1 > T \} \big| I_T \big)  \nonumber \\ 
\widetilde{J}_k^T(I_k) & = & \min \Big\{ P \big( \{ \Gamma_1 > k \} 
\big|I_k \big), {\hspace{0.06in}}
c P \big( \{ \Gamma_1 \leq k \} \big| I_k \big)  
+  E \left[ \widetilde{J}_{k+1}^T(I_{k+1}) \big| I_k \right] 
\Big \}, 
{\hspace{0.05in}} 0 \leq k < T  \nonumber 
\end{eqnarray}
where $I_0$ is the empty set. The first term in the above minimization 
corresponds to the cost associated with stopping at time $k$, while the 
second term corresponds to the cost associated with proceeding to time 
$k+1$ without stopping. The minimum expected cost for the finite-horizon 
optimization problem is $\widetilde{J}_0^T( I_0 )$.

\noindent {\bf \em Recursion for the Sufficient Statistics:} Consider 
the special case where change at all the sensors happens at 
the same instant. In this setting, it can be shown that the random 
variable $p_k \triangleq P \big( \{ \Gamma_1 \leq k \} \big| I_k \big)$ 
serves 
as the sufficient statistic for the above dynamic program and affords a 
recursion~\cite{venu_qdecentral}. To consider the more general case, 
we define an $(L+1)$-tuple of conditional probabilities, 
$\{ p_{k,\ell}, {\hspace{0.05in}} \ell = 1, \cdots, L+1  \}$: 
\begin{eqnarray} 
p_{k,\ell} & \triangleq &  
P\Big( \big\{ \Gamma_1 \leq k, \cdots, \Gamma_{\ell - 1} \leq k, 
\Gamma_{\ell } > k , \cdots, \Gamma_{L} > k \big \} \big| 
I_k \Big). 
\nonumber 
\end{eqnarray}
The special setting of~\cite{venu_qdecentral} is then equivalent to 
\begin{eqnarray}
p_{k, L+1} = p_k, {\hspace{0.1in}} p_{k,1} = 1 - p_k, 
{\hspace{0.1in}} {\rm and} {\hspace{0.1in}} 
p_{k,\ell} = 0, {\hspace{0.1in}} \ell = 2, \cdots, L. 
\nonumber 
\end{eqnarray} 

We now show that $\pb_k \triangleq [p_{k,1}, \cdots , p_{k,L+1} ]$ 
can be obtained from $\pb_{k-1}$ 
via a recursive approach. 
For this, we note that the underlying probability space $\Omega$ in the setup 
can be partitioned as 
\begin{eqnarray} 
\Omega & = & \bigcup_{\ell = 1}^{L+1} T_{k,\ell} \nonumber \\ 
T_{k,\ell} & \triangleq & 
\big\{ \Gamma_1 \leq k , \cdots , 
\Gamma_{\ell-1} \leq k, 
\Gamma_{\ell } \geq k+1 , 
\cdots , \Gamma_{L} \geq k+1 \big\}. \nonumber
\end{eqnarray} 
The event where no sensor has observed the change is denoted as $T_{k,1}$. 
(The test that will be proposed and studied later in the paper 
thresholds the {\em a posteriori} probability of $T_{k,1}$.) 
On the other hand, $T_{k,\ell}$ (for $\ell \geq 2$) 
corresponds to the event where the maximal index of the sensor that 
has observed the change before time instant $k$ is $\ell - 1$. 
Observe that $p_{k,\ell}$ is the probability of $T_{k,\ell}$ 
conditioned on $I_k$. 

To show that $p_{k,\ell}$ can be written in terms of 
$\pb_{k-1}$, the observations $\Zb_k$ and the prior 
probabilities, we partition $T_{k,\ell}$ further as 
\begin{eqnarray}
T_{k,\ell} & = & \bigcup_{j=1}^{\ell} U_{k,\ell,j} \nonumber \\ 
U_{k,\ell, j} & \triangleq 
& \big \{  \Gamma_1 \leq k-1, \cdots , 
\Gamma_{j-1} \leq k-1, 
\Gamma_{j} = k , \cdots, \Gamma_{\ell - 1} = k, \nonumber \\ 
&& {\hspace{0.1in}} \Gamma_{\ell} \geq k+1 , \cdots, \Gamma_{L} \geq k +1 
\big\}, {\hspace{0.05in}} 
1 \leq j \leq \ell. \nonumber 
\end{eqnarray}
Note that $U_{k,\ell, j } \cap T_{k-1, j} = U_{k,\ell, j}$. Using the new partition 
$\{ U_{k,\ell, j }, {\hspace{0.05in}} 
j = 1, \cdots, \ell \}$ and applying Bayes' rule repeatedly, it can be 
checked that $p_{k,\ell}$ can be written as 
\begin{eqnarray}
p_{k,\ell}  =  \frac{ 
\sum_{m = 1}^{\ell} 
f \left ( \Zb_k| I_{k-1}, U_{k,\ell, m}) 
P( U_{k,\ell, m} | I_{k-1}  \right)}
{ \sum_{j=1}^{L+1} \sum_{m=1}^j f( \Zb_k|I_{k-1}, U_{k,j,m} ) 
P( U_{k,j,m}|I_{k-1} )  } \triangleq 
\frac{ {\cal N}_{\ell} }{ \sum_{j = 1}^{L+1} {\cal N}_j } 
\nonumber  
\end{eqnarray}
where $f(\cdot | \cdot)$ denotes the conditional probability 
density function of $\Zb_k$ and ${\cal N}_{\ell}$ denotes the 
numerator term. 

From the i.i.d.\ assumption on the statistics of the observations, the 
first term within the summation for ${\cal N}_{\ell}$ can be written as: 
\begin{eqnarray}
f \left ( \Zb_k| I_{k-1}, U_{k,\ell, m} \right) = 
\prod_{j = 1}^{\ell - 1} f_1(Z_{k, j} ) \prod_{j = \ell }^L f_0(Z_{k,j}) 
= \prod_{j = 1}^{\ell - 1}L_{k,j} \prod_{j = 1}^L f_0(Z_{k,j}) 
\nonumber 
\end{eqnarray}
where $L_{k,j} \triangleq \frac{f_1(Z_{k,j} )}{f_0(Z_{k,j} )}$ is 
the likelihood ratio of the two hypotheses given that $Z_{k,j}$ is 
observed at the $j$-th sensor at the $k$-th instant. For the second 
term, observe from the definitions that 
\begin{eqnarray}
P(U_{k, \ell, m} | I_{k-1} ) = P(T_{k-1, m} | I_{k-1}) 
\frac{ P(U_{k, \ell, m}) } {P(T_{k-1, m })}. \nonumber 
\end{eqnarray} 
Thus, we have 
\begin{eqnarray}
{\cal N}_{\ell} & = & 
\left( \sum_{m = 1}^{\ell} \frac{ P(U_{k,\ell,m}) }  {P(T_{k-1, m})}  
\cdot p_{k-1, m} 
\right) 
\times 
\prod_{m = 1}^{\ell - 1}L_{k,m} \prod_{m = 1}^L f_0(Z_{k,m}) 
\nonumber \\ 
& \triangleq &  
\left( \sum_{m = 1}^{\ell} w_{k,\ell, m} {\hspace{0.03in}} p_{k-1, m} \right) 
{\bf \Phi}_{\sf obs}(k, \ell)  
\nonumber 
\end{eqnarray} 
where the first part is a weighted sum of $p_{k-1,m}$ with weights 
decided by the prior probabilities, and the second part of 
the evolution equation, ${\bf \Phi}_{\sf obs}(k,\ell)$, can be viewed as that 
part that depends only on the observation $\Zb_k$. 

Many observations are in order at this stage: 
\begin{itemize} 
\item 
The above expansion for ${\cal N}_{\ell}$ can be easily explained 
intuitively: If the maximal sensor index 
observing the change by time $k$ is $\ell-1$, then the maximal 
sensor index 
observing the change by time $k-1$ should be from the set 
$\{ 0, \cdots, \ell-1 \}$. 

\item 
Using the joint-geometric model for $\{ \Gamma_{\ell} \}$, it can 
be shown that $w_{k,\ell, m}$ is of the form: 
\begin{eqnarray}
w_{k,\ell, m } 
& = & \frac{ P(U_{k,\ell, m}) } {P(T_{k-1, m})} = 
\left( 1 - \rho_{\ell - 1,\ell} \right) \cdot 
\prod_{j = m-1 }^{\ell - 2} \rho_{j, j+1 }   \triangleq 
\left( 1 - \rho_{\ell - 1,\ell} \right) \cdot w_{m}^{\ell} 
\nonumber \\ 
{\cal N}_{\ell} & = & 
\prod_{m = 1}^{\ell - 1}L_{k,m} \prod_{m = 1}^L f_0(Z_{k,m}) 
\cdot \left( 1 - \rho_{\ell - 1, \ell } \right) 
\times \left( 
\sum_{m = 1}^{\ell} p_{k-1, m} \cdot w_{m}^{\ell}
\right) \label{one}
\end{eqnarray}
with the understanding that 
the product term in the definition of $w_m^{\ell}$ is vacuous (and is to 
be replaced by $1$) if $m = \ell$. It is important to note that the 
joint-geometric assumption renders the weights ($w_{k,\ell,m}$) associated 
with $p_{k-1,m}$ independent of $k$. This will be useful later in establishing 
convergence properties for the DP. 

\item 
It is important to note that given a fixed value of $\ell$, 
$p_{k,\ell}$ is dependent on the entire vector $\pb_{k-1}$ and not 
on $p_{k-1,\ell}$ alone. Thus, the recursion for ${\cal N}_{\ell}$ 
implies that $\pb_k$ forms the sufficient statistic and the 
function $\widetilde{J}_k^T(I_k)$ can be written as a function of 
only $\pb_k$, say $J_k^T( \pb_k)$. The finite-horizon DP equations 
can then be rewritten as 
\begin{eqnarray}
J_T^T( \pb_T) & = & p_{T,1} \nonumber \\ 
J_k^T( \pb_k ) & = & \min \Big\{  p_{k,1} , {\hspace{0.05in}} 
c(1 - p_{k,1})  
+ A_k^T( \pb_k ) \Big \} \nonumber 
\end{eqnarray}
with 
\begin{eqnarray}
A_k^T( \pb_k ) & \triangleq & E[ J_{k+1}^T(\pb_{k+1}) |I_k ] 
\nonumber \\ & = & 
\int \left[ J_{k+1}^T \left( \pb_{k+1} 
\right) f \left( \Zb_{k+1}|I_k \right) \right] \Big|_{ \Zb_{k+1} = \zb } d\zb. 
\nonumber 
\end{eqnarray}
The previously established recursion for $\pb_{k+1}$ ensures that the 
right-hand side is indeed a function of $\pb_k$. 


\item 
It is easy to check that the general framework reduces to the special case 
when all the change-points coincide with $\Gamma_1$~\cite{venu_qdecentral}. 
In this case, only $T_{k,1}$ and $T_{k,L+1}$ are non-empty sets with 
\begin{eqnarray}
T_{k,1} = \{ \Gamma_1 \geq k+1 \}, {\hspace{0.1in}} 
& {\rm and} & T_{k,L+1} = \{ \Gamma_1 \leq k \} ,
\nonumber \\ 
p_{k,L+1} = p_k, {\hspace{0.1in}} 
p_{k,1} = 1 - p_k  & {\rm and} & 
p_{k,\ell} = 0, {\hspace{0.1in}} \ell = 2, \cdots, L. 
\nonumber 
\end{eqnarray} 
Furthermore, the recursion for $p_k$ reduces to 
\begin{eqnarray}
p_k & = & \frac{{\cal N}} 
{  \prod_{j = 1}^L f_0 (Z_{k, j}) \left( 1 - p_{k-1} \right) (1 - \rho) + {\cal N} }
\nonumber \\ 
{\cal N} & = & 
\prod_{j = 1}^L f_1(Z_{k,j}) \left( (1 - p_{k-1})\rho + p_{k-1}  
\right) \nonumber 
\end{eqnarray}
which coincides with~\cite[eqn.\ (13)-(15)]{venu_qdecentral}. This case can also 
be obtained from the formula in~(\ref{one}) by setting $\rho_{\ell-1,\ell} = 1$ 
for all $\ell$ with $2 \leq \ell \leq L$. 
\end{itemize}


\section{Structure Of The Optimal Stopping Rule $(\tau_{\sf opt})$} 
\label{sec4}
The goal of this section is to study the structure of the optimal 
stopping rule, $\tau_{\sf opt}$. For this, we follow the same 
outline as in~\cite{venu_qdecentral,venu_sequential} (see, 
also~\cite[p.\ 133]{bertsekas} for a similar example) and study the 
infinite-horizon version of the DP problem by letting $T \rightarrow \infty$. 

\begin{theorem}
\label{thm_structure}
Let $\pb = [p_1, \cdots , p_{L+1}]$ be an element of the standard 
$L$-dimensional simplex ${\cal P}$, defined as, ${\cal P} \triangleq 
\big\{ \pb {\hspace{0.03in}} : {\hspace{0.03in}} \sum_{j=1}^{L+1} 
p_j = 1\big\}$. The infinite-horizon cost-to-go for the DP is of the form 
\begin{eqnarray}
J(\pb) = \min \Big\{ p_1,  {\hspace{0.05in}} 
c(1 - p_1) + A_J(\pb) \Big\}, 
\nonumber 
\end{eqnarray}
where the function $A_J(\pb)$: i) is concave in $\pb$ over ${\cal P}$; 
ii) is bounded as $0 \leq A_J(\pb) \leq 1$; 
and iii) satisfies $A_J(\pb) = 0$ over the hyperplane 
$\big\{\pb: p_1 = 0 \big\}$. 
\end{theorem}
\begin{proof}
Before considering the infinite-horizon DP, we will study the finite-horizon 
version and establish some properties along the directions 
of~\cite{venu_qdecentral,venu_sequential,bertsekas}. A 
straightforward induction argument 
shows that if $T$ is fixed, 
\begin{eqnarray}
0 \leq J_k^T(\pb) \leq 1 {\hspace{0.05in}} {\rm for} {\hspace{0.05in}}
{\rm all} {\hspace{0.05in}} 0 \leq k \leq T,  
\nonumber \\ 
0 \leq A_k^T(\pb) \leq 1 {\hspace{0.05in}} {\rm for} {\hspace{0.05in}}
{\rm all} {\hspace{0.05in}} 0 \leq k \leq T. \nonumber 
\end{eqnarray}
Similarly, it is easy to observe that for any $k$, 
$A_k^T(\pb)$ and $J_k^T(\pb)$ equal zero if $p_1 = 0$. In 
Appendix~\ref{app_ccv}, the concavity of 
$A_k^T(\cdot)$ and $J_k^T(\cdot)$ are established via a routine 
induction argument. 

We now consider the infinite-horizon DP and show that it is well-defined. 
(That is, we remove the restriction that the stopping time is finite and 
let $T \rightarrow \infty$.) Towards this end, we need to establish that 
$\lim \limits_{T} J_k^T(\cdot)$ exists, which is done as follows: By an 
induction argument, we note that for any $\pb$ and $T$ fixed, we have 
\begin{eqnarray}
J_k^T( \pb ) \leq J_{k+1}^T(\pb), {\hspace{0.05in}} 
0 \leq k \leq T-1. \nonumber 
\end{eqnarray}
It is important to note that this conclusion critically depends on the 
joint-geometric assumption of the change process (in particular, the 
{\em memorylessness} property that results in the independence of $w_{k,\ell,m}$ 
on $k$ in~(\ref{one})) and the i.i.d.\ nature of the observation process 
conditioned on the change-point. 

Using a similar induction approach, observe that 
for any $\pb$ and $k$ fixed, $J_k^{T+1}(\pb) \leq J_k^T(\pb)$. 
Heuristically, this can also be seen to be true because the 
set of stopping times increases with $T$. Since $J_k^T(\pb) \geq 0$ for 
all $k$ and $T$, for any fixed $k$, we can let $T \rightarrow \infty$ and we 
have 
\begin{eqnarray}
\lim_T J_k^T(\pb) = \inf_{T {\hspace{0.02in}} : {\hspace{0.02in}}
T {\hspace{0.02in}}  > {\hspace{0.02in}} k} 
J_k^T(\pb) \triangleq J_k^{\infty}(\pb). \nonumber 
\end{eqnarray} 
Furthermore, the {\em memorylessness} property and the i.i.d.\ observation 
process results in the invariance of $J_k^{\infty}(\pb)$ on $k$. This 
can be shown by a simple time-shift argument. Denote this common limit 
as $J(\pb)$. 

A simple dominated convergence argument~\cite{durrett} then shows that 
$\lim \limits_{T} A_k^T(\pb)$ is well-defined and independent of $k$. 
If we denote this limit as $A_J(\pb)$, we have 
\begin{eqnarray}
A_J(\pb) & = & \int \left[ J(\pb) f\big(\Zb \big| I_{\bullet} \big) 
\right] \Big|_{ \Zb = \zb } d \zb \nonumber \\ 
& = & \int J(\pb) \Big\{ \sum_{j=1}^{L+1} \big( 
(1 - \rho_{j-1,j}) \cdot 
\sum_{m = 1}^j w_{m}^j {\hspace{0.05in}} p_m \big) 
{\bf \Phi}_{ {\sf obs} } (\bullet, j) \Big \} \Big|_{ \Zb = \zb  } 
d \zb, \nonumber 
\end{eqnarray}
where the fact that ${\bf \Phi}_{ {\sf obs} }(k, j) \big|_{ \Zb = \zb }$ 
is independent of $k$ is denoted as ${\bf \Phi}_{ {\sf obs} }(\bullet, j)$. 
Hence, the infinite-horizon cost-to-go can be written as 
\begin{eqnarray}
J(\pb) = \min \Big \{ p_1, {\hspace{0.05in}} 
c(1 - p_1) + A_J(\pb) \Big \}. 
\nonumber 
\end{eqnarray}
The structure of $A_J(\pb)$ follows from the finite-horizon characterization 
by letting $T \rightarrow \infty$. 
\end{proof}

At this stage, it is a straightforward consequence that the optimal 
stopping rule is of the form 
\begin{eqnarray}
\tau_{\sf opt} = 
\inf_k \Big \{ p_{k,1}(1 + c) - c <  A_J(\pb_k) \Big \}. 
\nonumber 
\end{eqnarray}
That is, a change is declared when the hyperplane on the left side 
is exceeded by $A_J(\pb_k)$ and no change is declared, otherwise. 

We will next see that this test characterization reduces to a 
degenerate one as $\rho \rightarrow 0$. To establish this degeneracy 
result, along the lines of~\cite{venu_qdecentral}, we now define a 
one-to-one and invertible transformation\footnote{It is important to 
note that the transformation in~\cite{venu_qdecentral} can be generalized 
in more than one direction. For example, i) $q_{k,\ell} =
\frac{ \sum_{j= \ell+1}^{L+1} p_{k,j} }{\rho {\hspace{0.01in}} 
p_{k,1} }$, ii) $q_{k,\ell} = \frac{1 - p_{k,\ell} } { \rho 
{\hspace{0.01in}} p_{k,1}}$ etc.\ are consistent with the definition 
in~\cite{venu_qdecentral}. While these definitions of $q_{k,\ell}$ 
ensure that the structure of $\tau_{\sf opt}$ (as $\rho \rightarrow 0$) 
becomes simple, the recursion for $q_{k,\ell}$ (and hence, an 
understanding of the performance of the proposed test) becomes more 
complicated. We believe that the definition of $q_{k,\ell}$, as provided 
here, is the most natural generalization in the goal of understanding 
the performance of change process detection schemes.}, 
$\{ q_{k,\ell}, {\hspace{0.05in}} \ell = 
1, \cdots, L+1 \}$, as follows: 
\begin{eqnarray}
q_{k,\ell} = \frac{p_{k,\ell}}{\rho \hspp p_{k,1}}. \nonumber  
\end{eqnarray} 
The inverse transformation is given by: 
\begin{eqnarray}
p_{k,\ell} & = & \frac{q_{k,\ell}}{ \sum_{j=1}^{L+1} q_{k,j} }, 
{\hspace{0.05in}} \ell = 1, \cdots, L+1, 
\nonumber 
\end{eqnarray}
which is equivalent to 
\begin{eqnarray} 
p_{k,1} & = & \frac{1}{ 1 + \rho \sum_{j =2 }^{L+1} q_{k,j} } 
{\hspace{0.1in}} {\rm and } {\hspace{0.1in}} 
p_{k,\ell}  = \frac{\rho {\hspace{0.03in}} q_{k,\ell} }
{ 1 + \rho \sum_{j =2 }^{L+1} q_{k,j} } ,  {\hspace{0.05in}} 
\ell = 2, \cdots, L+1. \nonumber 
\end{eqnarray} 
We can write $q_{0,\ell}$ in terms of the priors as 
\begin{eqnarray} 
q_{0,1} = \frac{p_{0,1}} {\rho \hspp p_{0,1}} & = & \frac{1} {\rho}, 
\nonumber \\ 
q_{0,\ell}  =  \frac{ p_{0,\ell} }{ \rho \hspp p_{0,1} } & = & 
\frac{ 
P\big( \{ \Gamma_1 = \cdots = \Gamma_{\ell-1} = 0, \Gamma_{\ell} > 0 \} \big) } 
{\rho \hspp P\big( \{ \Gamma_1 > 0 \} \big) } \nonumber \\
& = & \frac{ \prod_{j = 0}^{\ell-2} \rho_{j,j+1} 
\left(1 - \rho_{\ell-1,\ell} \right)} { \rho\hspp(1-\rho) } 
, {\hspace{0.05in}} \ell = 2, \cdots, L+1. \nonumber
\end{eqnarray}
Note that while $p_{k,\ell}$ are conditional probabilities of certain 
events and hence lie in the interval $[0,1]$, the range of $q_{k,\ell}$ 
is in general $[0, \infty)$.

It can be checked that the evolution equation can be rewritten 
in terms of $q_{k,\ell}$ as 
\begin{eqnarray}
q_{k,\ell} = \frac{ 1 - \rho_{\ell-1,\ell} }{1 - \rho} 
\cdot \prod_{j=1}^{\ell -1} L_{k,j} \cdot \left( 
\sum_{j=1}^{\ell} q_{k-1, j} w_{j}^{\ell} \right). 
\label{r_kell_recursion} 
\end{eqnarray}
It is interesting to note from~(\ref{r_kell_recursion}) that the 
update for $q_{k,\ell}$ is a weighted sum of $q_{k-1,j}, j = 1, \cdots, \ell$ 
with progressively increasing weight as $j$ increases. Similarly, we can 
define $J_k^T(\cdot)$ and $A_k^T(\cdot)$ in terms of $\qb_k$. Using the 
transformation $\{ q_{k,\ell}\}$, $\tau_{\sf opt}$ is seen to have the form: 
{\vspace{0.05in}}
\begin{equation} 
\fbox{$\displaystyle 
\tau_{\sf opt} 
= \inf_k \left \{ 
\sum_{\ell = 2}^{L+1} q_{k,\ell} > \frac{1 - A_J (\qb_k) } 
{\rho \left( c + A_J(\qb_k)\right)} 
\right \}.$ \nonumber }
\end{equation}

{\vspace{0.1in}}
When all $\Gamma_{\ell}$ coincide~\cite{venu_qdecentral}, we have 
\begin{eqnarray}
q_{k, L+1} = \frac{p_k }{  \rho(1 - p_k)} \triangleq q_k, 
{\hspace{0.05in}} q_{k,1} = \frac{1}{\rho}, 
{\hspace{0.05in}} q_{k,\ell} = 0, {\hspace{0.05in}} 
\ell = 2, \cdots, L. \nonumber 
\end{eqnarray}
Further, it is straightforward to check that the evolution 
in~(\ref{r_kell_recursion}) reduces to 
\begin{eqnarray}
q_{k,L+1} = \frac{ \prod_{j= 1}^L L_{k,j} }{1 - \rho} \cdot 
\left( 1 + q_{k-1, L+1}  \right), 
\label{qk_rec}
\end{eqnarray}
which is~\cite[eqn.\ 32]{venu_qdecentral}. Thus, the space of sufficient 
statistics and the optimal test reduce to a one-dimensional variable 
($p_k = P \big( \{ \Gamma_1 \leq k \} \big | I_k \big)$ or equivalently, 
$q_k$) and a threshold test on $p_k$ (or equivalently, on $q_k$), respectively.

In the general case, unless something more is known about the structure 
of $A_J(\cdot)$ (which is possible if there is some structure on 
$\{ \rho_{\ell-1,\ell} \}$), we cannot say more about $\tau_{\sf opt }$. 
Nevertheless, the following theorem establishes its structure in the 
practical setting of a rare disruption regime ($\rho \rightarrow 0$). 
The limiting test thresholds the {\em a posteriori} probability that 
no-change has happened (from below), and is denoted as $\nu_{A  }$. 
\begin{theorem} 
\label{thm_rho0}
The structure of $\tau_{\sf opt}$ converges to a simple threshold 
rule 
in the asymptotic limit as $\rho \rightarrow 0$. This test 
is of the form: 
\begin{eqnarray}
\fbox {$ \displaystyle 
\nu_{A  }  = \left\{ 
\begin{array}{cc}
{\rm Stop} & {\rm if} {\hspace{0.05in}} 
\log \left( \sum_{\ell = 2}^{L+1} q_{k,\ell} \right) 
\geq A \\ 
{\rm Continue} & {\rm if} {\hspace{0.05in}} 
\log\left( \sum_{\ell = 2}^{L+1} q_{k,\ell}\right) < 
A 
\end{array} \nonumber 
\right. 
$}
\end{eqnarray}
{\vspace{0.1in}}
for an appropriate choice of threshold $A$. 
\end{theorem} 
\begin{proof} 
See Appendix~\ref{app_rho0}. 
\end{proof}

The test $\nu_{A  }$ is of low-complexity because of the following properties: 
i) a simple recursion formula~(\ref{r_kell_recursion}) for the sufficient 
statistics; ii) a threshold operation for stopping; and iii) the threshold 
value that can be pre-computed given the $P_{\sf FA}$ constraint (see 
Prop.~\ref{prop_pfa}). However, it is important to note that the complexity 
of $\nu_{A  }$ is {\em not} equivalent to that of the threshold test 
of~\cite{venu_qdecentral} because the recursion for the sufficient statistics 
depends on $(L+1)$ {\em a posteriori} probabilities, in general, in contrast to 
a single parameter in~\cite{venu_qdecentral}. 

The fact that $\tau_{\sf opt} \stackrel {\rho \downarrow 0}{\rightarrow} 
\nu_{A  }$ for an appropriate choice of $A$ {\em does not} imply that 
$\nu_{A  }$ is 
asymptotically (as $\rho \rightarrow 0$ or as $P_{\sf FA} \rightarrow 0$) 
optimal. However, the low-complexity of this test, in addition to 
Theorem~\ref{thm_rho0}, and the fact that the structure of $A_J(\qb_k)$ 
(and hence, $\tau_{\sf opt}$) are not known suggest that it is a good 
candidate test for change detection across a sensor array. 
In fact, we will see this to be the case when we establish sufficient 
conditions under which $\nu_{A  }$ is asymptotically optimal.

\section{Main Results on $\nu_A$} 
\label{sectionx} 
Towards this end, our main interest is in understanding the performance 
($E_{\sf DD}$ vs.\ $P_{\sf FA}$) of $\nu_{A  }$ for any general choice of 
threshold $A$. We make a few preliminary remarks before providing 
performance bounds for $\nu_A$. 

\noindent {\bf \em Special Cases of Change Parameters:} We start by 
considering some special scenarios of change propagation modeling. 
The first scenario corresponds to the case where one (or more) of the 
$\rho_{\ell-1,\ell}$ is $1$. The following proposition addresses this 
setting. 
\begin{prop}
\label{prop_oblivious}
Consider an $L$-sensor system described in Sec.~\ref{sec2}, 
parameterized by $\{ \rho_{\ell-1,\ell} \}$, where $\rho_{\ell', \ell'+1} 
= 1$ for some $\ell'$ and $\max \limits_{j \neq \ell'} \rho_{j, j+1} < 1$. 
This system is equivalent to an $(L-1)$-sensor system, parameterized by 
$\{ \beta_{\ell, \ell+1}  \}$, where 
\begin{eqnarray}
\beta_{j, j+1} & = & 
\rho_{j, j+1}, {\hspace{0.05in}} j \leq \ell' - 1 
\nonumber \\ 
\beta_{j, j+1} & = & \rho_{j+1, j+2}, {\hspace{0.05in}} 
j \geq \ell' \nonumber
\end{eqnarray}
with the $(\ell'+1)$-th sensor observing (a combination of) 
$Z_{k, \ell'+1}$ and $Z_{k, \ell'+2}$ with a geometric delay 
parameter of $\beta_{\ell', \ell'+1} = \rho_{\ell'+1, \ell'+2}$. 
\end{prop}
\begin{proof}
The proof is straightforward by studying the evolution of 
$\{ q_{k,\ell} \}$ for the original $L$-sensor system. 
From~(\ref{r_kell_recursion}), it can be seen that $q_{k,\ell'+1} = 0$ 
(identically) for all $k$ and the reduced $(L-1)$-dimensional system 
discards this redundant information, while the observation corresponding 
to the $(\ell'+1)$-th sensor is carried over to the $(\ell'+2)$-th original 
sensor. 
\end{proof}

The second scenario corresponds to the case where one (or more) 
of the $\rho_{\ell-1,\ell}$ is $0$. 
\begin{prop}
\label{prop_blocking}
Consider an $L$-sensor system, parameterized by $\{ \rho_{\ell-1,\ell} \}$, 
with $\ell'$ indicating the smallest index such that 
$\rho_{\ell', \ell'+1} = 0$. This system is equivalent to an 
$\ell'$-sensor system with the same parameters as that of the original 
system. It is as if sensors $(\ell'+1)$ and beyond do not exist (or 
contribute) in the context of change detection. 
\end{prop}
\begin{proof}
The proof is again straightforward by considering the evolution of 
$\{ q_{k,\ell} \}$ in~(\ref{r_kell_recursion}) and noting that 
$q_{k,j}, {\hspace{0.05in}} j \geq \ell'+2$ are identically $0$ for 
all $k$. 
\end{proof}

It is useful to interpret Props.~\ref{prop_oblivious} 
and~\ref{prop_blocking} via an ``information flow'' paradigm. If 
change propagation is instantaneous across a sensor (corresponding to 
the first case), it is as if the fusion center is {\em oblivious} to 
the presence of that sensor conditioned upon the previous sensors' 
observations. In this setting, the detection delay corresponding to 
that sensor is zero, as would be expected from the fact that the 
geometric parameter is $1$. In the second case, information flow to 
the fusion center (concerning change) is {\em cut-off or blocked} past 
the first sensor with a geometric parameter of $0$. That is, the 
observations made by sensors $\{\ell'+1, \cdots, L  \}$ (if any) do not 
contribute information to the fusion center in helping it decide whether 
the disruption has happened or not. Apart from these extreme cases of 
oblivious$/$blocking sensors, we can assume without 
any loss in generality that 
\begin{eqnarray} 
0 < \min \limits_{\ell} \rho_{\ell-1,\ell} \leq 
\max \limits _{\ell} \rho_{\ell-1,\ell} < 1. \nonumber 
\end{eqnarray}
Continuity arguments suggest that if some $\rho_{\ell-1,\ell}$ is small 
(but non-zero), it should be natural to expect that the $\ell$-th sensor 
and beyond {\em may not} ``effectively'' contribute any information to 
the fusion center. We will interpret this observation after establishing 
tractable performance bounds for $\nu_A$.

\noindent{\bf \em Probability of False Alarm:} We first show that letting 
$A \rightarrow \infty$ in $\nu_A$ corresponds to considering the regime 
where $P_{\sf FA} \rightarrow 0$. 
\begin{prop}
\label{prop_pfa}
The probability of false alarm with $\nu_{ A}$ can be upper bounded as 
\begin{eqnarray}
P_{\sf FA} \leq \frac{1}{1 + \rho \cdot \exp(A)} 
.\nonumber 
\end{eqnarray}
That is, if $\alpha \leq 1$ and the threshold $A$ is set as $A = 
\log \left(\frac{1}{\rho \alpha} \right)$, then $P_{\sf FA} \leq \alpha$. 
\end{prop}
{\vspace{0.05in}}
\begin{proof}
The proof is elementary and follows the same argument as 
in~\cite{TVAS,venu_bt}. Note that $p_{k,1}$ and $\nu_{ A }$ 
can also be written as 
\begin{eqnarray}
p_{k,1} & = & P \big( \{ \Gamma_1 > k  \} \big| I_k \big) \nonumber \\ 
\nu_{  A } &= & \inf_k \left\{  p_{k,1} \leq 
\frac{1}{ 1 + \rho \cdot\exp(A)} 
\right\}. \nonumber 
\end{eqnarray}
Thus, we have 
\begin{eqnarray}
P_{\sf FA} = P \big( \{ \nu_A < \Gamma_1 \} \big) 
= E \left[ p_{ \nu_A , 1  }  \right] 
\leq \frac{1}{1 + \rho \cdot \exp(A)}. \nonumber
\end{eqnarray}
\end{proof}

\noindent {\bf \em Universal Lower Bound on $E_{\sf DD}$:} We now 
establish a lower bound on $E_{\sf DD}$ for the class of stopping times 
${\bf \Delta}_{\alpha}$. That is, any stopping time $\tau$ should have 
an $E_{\sf DD}$ larger than the lower bound if $P_{\sf FA}$ is to be 
smaller than $\alpha$. 
\begin{prop}
\label{prop_init} 
Consider the class of stopping times ${\bf \Delta}_{\alpha} = \{  
\tau : P_{\sf FA}(\tau ) \leq \alpha \}$. Under the assumption that 
$\min \limits_{\ell = 2, \cdots, L} \rho_{\ell-1,\ell} > 0$, as 
$\alpha \rightarrow 0$, we have 
\begin{eqnarray}
\inf_{ \tau {\hspace{0.02in}} \in 
{\hspace{0.02in}} {\bf \Delta}_{\alpha} } 
E_{\sf DD}(\tau) 
\geq \frac{ \log \Big( \frac{1}{\rho \alpha} \Big) 
\cdot \left( 1 + {\rm o}(1) \right)} 
{ L D(f_1, f_0) + | \log(1 - \rho) |   } .  
\nonumber 
\end{eqnarray}
\end{prop}
{\vspace{0.15in}}
\begin{proof}
The proof follows on similar lines as~\cite[Lemma 1 and 
Theorem 1]{TVAS}, but with some modifications to accommodate 
the change process setup. See Appendix~\ref{app_prop_init}. 
\end{proof}

\noindent{\bf \em Upper Bound on $E_{\sf DD}$ of $\nu_A$:} We will 
establish an 
upper bound on $E_{\sf DD}$ of $\nu_A$. Using this bound, it can be seen 
that $\nu_{A}$ meets the lower bound (proved above) for an 
appropriate choice of $A$, thus establishing its asymptotic 
optimality. The main result is as follows. 
\begin{theorem} 
\label{thm_final}
Let $\{\rho_{\ell - 1, \ell} \}$ be such that $0 < \min \limits_{\ell } 
\rho_{\ell - 1, \ell} \leq \max \limits_{\ell} \rho_{\ell - 1,\ell} < 1$. 
Further, assume that $D(f_1, f_0)$ be such that there exists some $j$ 
satisfying $\ell \leq j \leq L$ and 
\begin{eqnarray} 
D(f_1, f_0) > \frac{1}{j - \ell +1} \log \left( \frac{\sum_{p = 0} ^{\ell - 1} 
(1 - \rho_{p, p+1})} {1 - \rho_{j, j+1}} \right), 
\label{cond_five}
\end{eqnarray}
for all $2 \leq \ell \leq L$. 
Then, $\nu_{A}$ with $A = \log \left( \frac{1}{\rho \alpha} \right)$ 
is asymptotically optimal (as $\alpha \rightarrow 0$). Furthermore, the 
performance of $\nu_{A}$ in this regime 
is of the form: 
{\vspace{0.05in}}
\begin{eqnarray} 
\fbox{$\displaystyle 
E_{\sf DD}
= \frac{ \log \left( \frac{1}{\rho }\right) + 
| \log(P_{\sf FA})|  
}{ L D(f_1, f_0) + 
|\log \left( 1 - \rho \right)| } + {\mathrm{o}}(1) . 
$ \nonumber } 
\end{eqnarray} 
\endproof 
\end{theorem}

The proof of Theorem~\ref{thm_final} in the general case of an arbitrary 
number ($L$) of sensors with an arbitrary choice of $\{ \rho_{\ell-1,\ell} \}$ 
results in cumbersome analysis. Hence, it is worthwhile considering the 
special case of two sensors that can be captured by just two change 
parameters: $\rho$ and $\rho_{1,2}$. The main idea that is necessary in 
tackling the general case is easily exposed in the $L = 2$ setting in 
Sec.~\ref{sectionxx}. The general case is subsequently studied in 
Sec.~\ref{sectionxxx}.

\section{Expected Detection Delay: Special Case ($L = 2$)} 
\label{sectionxx} 
The main statement in the $L = 2$ case is the following result. 

{\em Theorem 3 ($L = 2$):} The stopping time $\nu_A$ is such that 
$\nu_A \rightarrow \infty$ as $A \rightarrow \infty$. Further, if 
$D(f_1, f_0)$ satisfies 
\begin{eqnarray}
D(f_1, f_0) > \log \left( 2 - \rho - \rho_{1,2}   \right), \nonumber
\end{eqnarray}
as $A \rightarrow \infty$, we also have 
\begin{eqnarray}
E_{\sf DD} = E[\nu_A] 
\leq \frac{A}{ 2D(f_1, f_0) + |\log \left( 1 - \rho \right)| }. 
\nonumber 
\end{eqnarray}
\endproof 

We will work our way to the proof of the above statement by establishing 
some initial results. 
\begin{prop}
\label{prop_reduce_pre}
If $0 < \{ \rho, \rho_{1,2} \} < 1$, we can recast $\{q_{k,\ell} \}$ 
as follows: 
\begin{eqnarray}
q_{k,1} & = & \frac{1}{\rho} \nonumber \\
q_{k,2} & = & \underbrace{ 
\left ( \frac{1 - \rho_{1,2} } {1 - \rho} \right)^k \cdot 
\left(1 + \frac{1 - \rho_{1,2} }{1 - \rho}   \right) }_{ \alpha_{k,2} } 
\cdot 
\underbrace{ \prod_{m = 1}^k L_{m,1} }_{ C_1 } \cdot 
\underbrace{ \prod_{m = 0}^{k-2} \left( 1 + \zeta_{m,2} \right) }_{ J_2} 
\nonumber \\ 
\zeta_{m,2} & = & \frac{1 - \rho} { (1 - \rho_{1,2} ) \cdot 
(1 + q_{m,2} ) \cdot L_{m+1,1} } \nonumber \\ 
q_{k,3} & = & \underbrace{ 
\frac{\rho_{1,2} }{ \left(1 - \rho \right)^k} \cdot 
\left( 1 + \frac{1 - \rho_{1,2} }{1 - \rho} + \frac{1}{1 - \rho} 
\right) }_{ \alpha_{k,3} } 
\cdot \underbrace{ \prod_{m = 1}^k L_{m,1} L_{m,2} }_{C_1 C_2} 
\cdot \underbrace{ \prod_{m = 0}^{k-2} \left(1 + \zeta_{m,3} \right) 
}_{ J_3 } \nonumber \\ 
\zeta_{m,3} & = & \frac{  \rho_{1,2} \cdot 
\Big( 1 - \rho + (1 - \rho_{1,2} ) \cdot L_{m+1,1} \cdot 
(1 + q_{m,2} )  \Big) } 
{ L_{m+1,1} L_{m+1,2} \cdot 
\left( \rho_{1,2} + \rho_{1,2} \hspp q_{m,2} + q_{m,3}  \right) }. 
\nonumber 
\end{eqnarray}
\end{prop}
{\vspace{0.09in}}
\begin{proof}
We start with the recursions 
\begin{eqnarray}
q_{k,2} & = & \frac{ (1 - \rho_{1,2}) }{1 - \rho} \cdot L_{k,1} \cdot 
\left(1  + q_{k-1,2} \right) \nonumber \\ 
q_{k,3} & = & \frac{L_{k,1} L_{k,2} }{1 - \rho} \cdot \left( 
\rho_{1,2} + \rho_{1,2} \hspp q_{k-1,2} + q_{k-1,3} \right). \nonumber 
\end{eqnarray} 
The expression for $q_{k,2}$ is obtained by isolating the term 
$(1 + q_{k-j,2})$ at every stage as $j$ increases from $2$ to $k$. 
The expression for $q_{k,3}$ is obtained by isolating the term 
$\left( \rho_{1,2} + \rho_{1,2} \hspp q_{k-j,2}+ q_{k-j,3} \right)$ at every 
stage as $j$ increases. 
\end{proof}
The test $\nu_A$ can now be rewritten as 
\begin{eqnarray}
\nu_A & = & \inf \limits_k \Big\{ \log \left( q_{k,2} + q_{k,3}
\right) > A \Big\} \nonumber \\ 
& = & \inf \limits_k \Big\{ \log \left( \alpha_{k,2} \cdot C_1 \cdot J_2 + 
\alpha_{k,3} \cdot C_1 C_2 \cdot J_3 \right) > A \Big\}  \nonumber \\ 
& = & \inf \limits_k \bigg\{ \log(\alpha_{k,2} \cdot C_1 \cdot J_2) 
+ \log \left(  1 +  C_2 \cdot \frac{ \alpha_{k,3} } { \alpha_{k,2} } 
\cdot \frac{J_3}{J_2 }  \right)  
> A \bigg \}.  \nonumber 
\end{eqnarray}

We need the following preliminaries in the course of our analysis. 
\begin{lemma}
\label{lem_bd_pre}
Since $q_{m,2} \geq 0$, note that $J_2$ 
can be trivially upper bounded as 
\begin{eqnarray}
J_2 & \leq & \prod_{m = 1}^{k-1} \left( 1 + \frac{1 - \rho}
{(1 - \rho_{1,2} ) \cdot L_{m,1}}  \right). \nonumber 
\end{eqnarray}
\endproof 
\end{lemma}
\begin{lemma}
\label{lem_positive}
If $\{ x, x_1, x_2, \cdots \}$ are 
i.i.d.\ with $x \geq 0$ and $E[ \log(x) ] > 0$, then 
\begin{eqnarray} 
\frac{1}{k} \log \left( 1 + \prod_{m=1}^k x_m \right) 
- \frac{\sum _{m = 1}^k \log(x_m) }{k} 
\stackrel{k \rightarrow \infty} {\rightarrow} \hspp \hspp 0 \hspp  
{\hspace{0.05in}} {\rm a.s.\ } {\hspace{0.02in}} {\rm and } 
{\hspace{0.05in}} {\rm in} {\hspace{0.05in}} {\rm mean}. 
\nonumber 
\end{eqnarray}
If $\{x, x_1, x_2, \cdots  \}$ are 
i.i.d.\ with $x \geq 0$ and $E[ \log(x) ] \leq 0$, 
then 
\begin{eqnarray} 
\frac{1}{k} \log \left( 1 + \prod_{m=1}^k x_m \right) 
\stackrel{k \rightarrow \infty} {\rightarrow}
0 {\hspace{0.05in}} {\rm a.s.\ } {\hspace{0.02in}} {\rm and } 
{\hspace{0.05in}} {\rm in} {\hspace{0.05in}} {\rm mean}. 
\nonumber 
\end{eqnarray}
Note that both these conclusions are true even if $\{x_m \}$ are not 
i.i.d.\ (or even independent) as long as the condition on the sign of 
$E[ \log(x)]$ can be replaced with an almost sure (and in mean) 
statement on the sign of $\lim \limits_n \frac{1}{n}\sum_{m = 1 }^n \log( x_m )$ 
(or an appropriate variant thereof). 
\endproof 
\end{lemma}
The following statement, commonly referred to as the 
Blackwell's elementary renewal theorem~\cite[pp.\ 204-205]{durrett}, 
is needed in our proofs.
\begin{lemma} 
\label{lem_blackwell}
Let $x_m$ be i.i.d.\ positive random variables and define $T_m$ as 
follows: 
\begin{eqnarray} 
T_m = T_{m-1} + x_m, {\hspace{0.05in}} m \geq 1 
{\hspace{0.05in}} {\rm and} {\hspace{0.05in}} T_0 = 0. 
\nonumber 
\end{eqnarray}
The number of renewals in $[0, t]$ is 
$N_t = \inf \limits_k \Big\{ T_k > t \Big\}$. Then, we have 
\begin{eqnarray} 
\frac{N_t }{t} & \rightarrow & \frac{1}{\mu} {\hspace{0.08in}} {\rm a.s.} 
{\hspace{0.08in}} {\rm as} {\hspace{0.08in}} t \rightarrow \infty 
{\hspace{0.08in}} {\rm and} 
\nonumber \\ 
\frac{ E[ N_t ] }{t} & \rightarrow & \frac{1}{\mu} {\hspace{0.08in}} 
{\rm as} {\hspace{0.08in}} t \rightarrow \infty, \nonumber 
\end{eqnarray}
where $\mu \triangleq E[x_m] \in (0, \infty]$. \endproof 
\end{lemma}

\noindent {\bf \em Proof of Theorem~\ref{thm_final} ($L = 2$):} We will 
postpone the 
proof of the first statement to Sec.~\ref{sectionxxx} when we consider the 
general case in Prop.~\ref{prop_infty_bd}. For the second statement, we 
first use the bound for $J_2$ from Lemma~\ref{lem_bd_pre} and the fact that 
$\zeta_{m,\ell} \geq 0$, and thus we have 
\begin{eqnarray}
\log \left( 1 +  C_2 \cdot \frac{ \alpha_{k,3} } { \alpha_{k,2} } 
\cdot \frac{J_3}{J_2 } \right) & \geq &
\log \left( 1 +  C_2 \cdot \frac{ \alpha_{k,3} } { \alpha_{k,2} } 
\cdot \frac{1}{ \prod_{m=1}^{k-1} \left(1 
+ \frac{1 - \rho}{ (1 - \rho_{1,2} )L_{m,1} }  \right)} \right) 
\nonumber \\ 
& \geq & \log \left(1 + \prod_{m=1}^k 
\frac{ \rho_{1,2}^{1/k} \cdot L_{m,2} }
{ \left( 1 - \rho_{1,2} \right) \cdot 
\left(1 + \frac{1 - \rho}{ (1 - \rho_{1,2} )L_{m,1} }  \right) }  
\right). 
\nonumber 
\end{eqnarray}
Now, observe that 
\begin{equation} 
\begin{split}
& E \left[ \log \left(  \frac{ L_{m,2} }
{ \left( 1 - \rho_{1,2} \right) \cdot 
\left(1 + \frac{1 - \rho}{ (1 - \rho_{1,2} )L_{m,1} }  \right) }
\right)  \right] \nonumber \\ 
& \quad \quad \quad =  D(f_1, f_0) + \log \left( \frac{1}{1 - \rho_{1,2}} \right) 
- E \left[ \log \left( 1 + \frac{1 - \rho}{ (1 - \rho_{1,2} )L_{m,1} }
\right) \right] \nonumber \\ 
& \quad \quad \quad \geq  D(f_1, f_0) 
+ \log \left( \frac{1}{1 - \rho_{1,2}} \right) 
-  \log \left( 1 + E \left[ 
\frac{1 - \rho}{ (1 - \rho_{1,2} )L_{m,1} } \right] \right) \nonumber \\ 
& \quad \quad \quad 
=  D(f_1, f_0) - \log \left(2 - \rho - \rho_{1,2}  \right) > 0 \nonumber
\end{split}
\end{equation}
where the first equality follows since $\rho_{1,2} > 0$ (change has to 
eventually happen at the second sensor to ensure that $E[\log (L_{m,2} )] = 
D(f_1, f_0)$), the second step 
follows from Jensen's inequality and the third equality from the fact that 
$E_{ f_1} \left[ \frac{1}{L_{m,1}} \right] = 1$. Using this fact in conjunction 
with Lemma~\ref{lem_positive} and noting that $\rho_{1,2} > 0$, as 
$k \rightarrow \infty$, we have 
\begin{eqnarray}
\log(\alpha_{k,2} \cdot C_1 \cdot J_2) 
+ \log \left(  1 +  C_2 \cdot \frac{ \alpha_{k,3} } { \alpha_{k,2} } 
\cdot \frac{J_3}{J_2 }  \right) & \geq & 
\log \left( C_1 C_2 \cdot \alpha_{k,3} 
\cdot J_3 \right) \nonumber \\ 
& \geq & \underbrace{ 
\sum_{m = 1}^k \log \left(  
\frac{ \rho_{1,2}^{1/k} \cdot L_{m,1} \cdot L_{m,2} }
{1 - \rho}   \right). }_{L_k} \nonumber 
\end{eqnarray}
The above relationship implies that $\nu_{A } \leq \nu_{L,A}$ where 
\begin{eqnarray}
\nu_{L,A} \triangleq  \inf_k \Big \{ L_k >A \Big \}. 
\nonumber
\end{eqnarray}
Applying Lemma~\ref{lem_blackwell} (since the entries in the definition of 
$\nu_{L,A}$ are independent) and the first statement of the theorem that 
$\nu_{A} \rightarrow \infty$ as $A \rightarrow \infty$, we have 
\begin{eqnarray}
\frac{E[\nu_A]}{A} \leq \frac{ E[\nu_{L,A}] }{A} 
\stackrel{A \rightarrow \infty}{\rightarrow } \frac{1}{2 D(f_1, f_0) + 
|\log \left( 1- \rho \right)| }. \nonumber 
\end{eqnarray}
\endproof

\section{Expected Detection Delay: General Case ($L \geq 3$)} 
\label{sectionxxx} 
We now consider the general case where $L \geq 3$. The main statement here 
is as follows. 

{\em Theorem 3 ($L \geq 3$):} If $D(f_1, f_0)$ is such that the 
condition~(\ref{cond_five}) is satisfied, as $A \rightarrow \infty$, 
we have 
\begin{eqnarray}
E_{\sf DD} = E[\nu_A] \leq \frac{A}{LD(f_1, f_0) + |\log\left(1 - \rho 
\right)|}. \nonumber
\end{eqnarray}
\endproof 

As before, we will work towards the proof of this statement. For this, 
the following generalizations of Prop.~\ref{prop_reduce_pre} and 
Lemma~\ref{lem_bd_pre} are necessary. 
\begin{prop}
\label{prop_reduce}
We have 
\begin{eqnarray}
q_{k,\ell} & = & \alpha_{k,\ell}
\cdot \prod_{j=1}^{\ell-1} \underbrace{ 
\prod_{m=1}^k  L_{m,j}  }_{C_j }
\cdot \underbrace{ \prod_{m = 0}^{k-2} 
\left( 1 + \zeta_{m,\ell} \right) }_{J_{\ell}} 
, {\hspace{0.07in}} \ell = 2, \cdots, L+1 
{\hspace{0.07in}} {\rm where} \nonumber \\ 
\alpha_{k,2} & = & \left( \frac{ 1 - \rho_{1,2} }{1 - \rho} \right)^k 
\cdot \left(1 + \frac{1 - \rho_{1,2}}{1- \rho}  \right) 
\nonumber \\ 
\alpha_{k,\ell} & = & 
\left( \frac{ 1-\rho_{\ell-1,\ell} }{1-\rho} \right)^{k} 
\cdot \prod_{j=1}^{\ell-2} \rho_{j,j+1}
\cdot \left ( \sum_{j=0}^{\ell-1} 
\frac{1 - \rho_{j,j+1}}{1 - \rho} \right), {\hspace{0.05in}} 
\ell \geq 3
\nonumber \\ 
\zeta_{m,\ell} & = & 
\frac{1}{(1 - \rho_{\ell-1,\ell}) \cdot \prod_{j=1}^{\ell-1} L_{m+1, j} } 
\cdot 
\frac{ \sum_{j=1}^{\ell-1} q_{m,j} \hspp w_{j}^{\ell} \hspp C_{m+1, j,\ell}  } 
{ \sum_{j=1}^{\ell} q_{m,j} \hspp w_j^{\ell} } \nonumber \\ 
B_{m, n,\ell} & = & \sum_{p = n-1}^{\ell-1} (1 - \rho_{p,p+1} ) \cdot 
\prod_{j = 1}^p L_{m,j}, {\hspace{0.05in}} 
n = 1, \cdots, \ell \nonumber \\ 
C_{m, n,\ell} & = & B_{m, n,\ell} - (1 - \rho_{\ell-1,\ell} ) \cdot 
\prod_{j=1}^{\ell-1}L_{m,j}, {\hspace{0.05in}} n = 1, \cdots, \ell . 
\nonumber 
\end{eqnarray}
\end{prop}
{\vspace{0.05in}}
\begin{proof} 
The proof is provided in Appendix~\ref{app_completionx} for the sake of 
completeness. Also, see Appendix~\ref{app_completionx} for how this proposition 
can be reduced to the case of~\cite{venu_qdecentral}. 
\end{proof}
\begin{lemma}
\label{lem_bd}
The following upper bound for $\zeta_{m,\ell}$ is obvious when 
$\max \limits_{\ell} \rho_{\ell-1, \ell} < 1$: 
\begin{eqnarray}
\zeta_{m,\ell} \leq \frac{B_{m+1, 1,\ell} }
{ (1 - \rho_{\ell-1,\ell}) \cdot \prod_{j=1}^{\ell-1} L_{m+1, j} }
= \frac{ 
\sum_{p = 0}^{\ell-2} (1 - \rho_{p,p+1}) \prod_{j=1}^p L_{m+1, j} }
{ (1 - \rho_{\ell-1,\ell}) 
\cdot \prod_{j=1}^{\ell-1}L_{m+1, j} } . 
\nonumber
\end{eqnarray} 
\endproof 
\end{lemma}

From Prop.~\ref{prop_reduce}, $\nu_A$ 
can be conveniently rewritten as 
\begin{eqnarray}
\fbox{ $\displaystyle 
\nu_A  =  \inf_k \left\{  \log \left( \sum_{\ell=2}^{L+1} 
\alpha_{k,\ell} \cdot C_1 \cdots C_{\ell-1} \cdot J_{\ell} 
\right) > A \right\}. \nonumber $} 
\end{eqnarray}

Unlike the setting in Sec.~\ref{sectionxx}, the structure of $\nu_{ A }$ 
(as of now) is not amenable to studying $E_{\sf DD}$ (in further detail). 
This is because it has the form of log of sum of random variables 
(see~\cite{venu_bt} for similar difficulties in the multi-hypothesis 
testing problem). We alleviate this difficulty by rewriting the test 
statistic in terms of quantities whose asymptotics can be easily studied. 
\begin{prop} 
\label{lemma_test_stat} 
We have the following expansion for the test statistic: 
\begin{eqnarray} 
\log \left( \sum_{\ell =2}^{L+1} \alpha_{k,\ell} \cdot 
C_1 \cdots C_{\ell-1} \cdot J_{\ell} \right) & = & 
\log \left( \alpha_{k,2} \cdot C_1 \cdot J_2  \right) + 
\sum_{\ell=2}^{L} 
\log \left( 1 + \frac{ \eta_{\ell} \cdot \alpha_{k,\ell+1} 
\cdot C_{\ell} \cdot J_{\ell+1} }{ \alpha_{k,\ell} \cdot J_{\ell} } 
\right) \nonumber \\ 
& = & \log \left( \left( \frac{1 - \rho_{1,2} }{1- \rho} \right)^k 
\cdot \frac{2 - \rho - \rho_{1,2} }{1 - \rho} 
\cdot C_1 \cdot J_2  \right) \nonumber \\ 
& & {\hspace{0.1in}} + 
\sum_{\ell = 2}^L \log \left( 1 + \eta_{\ell} \cdot 
\beta_{k,\ell} \cdot C_{\ell} \cdot \frac{J_{\ell + 1}}{J_{\ell}} \right)
\nonumber 
\end{eqnarray}
where 
\begin{eqnarray}
\beta_{k,\ell} & = & \frac{\alpha_{k,\ell+1}}{  \alpha_{k,\ell}} = 
\left( \frac{1 - \rho_{\ell, \ell+ 1}} {1 - \rho_{\ell-1,\ell} } \right)^k 
\cdot \rho_{\ell-1,\ell} \cdot 
\left(  1 + \frac{ 1 - \rho_{\ell, \ell +1} } { 
\sum_{m = 0} ^{\ell - 1} 1 - \rho_{m, m+1}  } \right), 
{\hspace{0.1in}}
\ell = 2, \cdots, L  
\nonumber \\ 
\eta_{\ell+1} & =  & \frac{ \eta_{\ell} \cdot \beta_{k,\ell} \cdot 
C_{\ell} \cdot \frac{ J_{\ell+1} }{ J_{\ell} } } 
{ 1 +  
\eta_{\ell} \cdot \beta_{k,\ell} \cdot 
C_{\ell} \cdot \frac{ J_{\ell+1} }{ J_{\ell} }
}, 
{\hspace{0.1in}} \ell = 2, \cdots, L-1 
\nonumber 
\end{eqnarray}
with $\eta_2 = 1$. 
\end{prop} 
\begin{proof}
The proof is straightforward by using the induction principle. 
\end{proof}

The following proposition establishes the general asymptotic trend of $\nu_A$. 
\begin{prop}
\label{prop_infty_bd}
The test $\nu_A$ is such that $\nu_{ A } \rightarrow \infty$ a.s.\ 
as $A \rightarrow \infty$. 
\end{prop}
\begin{proof}
See Appendix~\ref{app_completionx}. 
\end{proof}

As we try to understand $\nu_A$ further, it is important to note that the 
behavior of the decision statistic of $\nu_A$ is determined (only) by the 
trends of 
\begin{eqnarray}
x_{\ell} \triangleq \beta_{k,\ell} \cdot C_{\ell} 
\cdot \frac{ J_{\ell+1} }{J_{\ell}}, 
{\hspace{0.1in}} \ell = 2, \cdots, L. 
\nonumber 
\end{eqnarray} 
This is so because the asymptotics of $\{ \eta_{\ell} \}$ are also 
primarily determined by the trends of $\{ x_{\ell} \}$. 
We now develop the generalized version of the heuristic in 
Sec.~\ref{sectionxx} for the upper bound of $E_{\sf DD}$. Consider the 
case where $L = 4$. The second piece in the description of the test 
statistic (in Prop.~\ref{lemma_test_stat}) can be written as 
\begin{eqnarray}
{\cal L} \triangleq  
\log \left(1 + \eta_2 x_2  \right) + \log \left( 1 + \eta_3 x_3 \right) 
+ \log \left(1 + \eta_4 x_4  \right ) \nonumber 
\end{eqnarray} 
where the evolution of $\eta_{\ell}$ and $x_{\ell}, {\hspace{0.05in}} 
\ell = 2,3, 4$ is described in Prop.~\ref{lemma_test_stat}. In the regime 
where $k \rightarrow \infty$, note that if $x_{2} \rightarrow \infty$ 
(with high probability), then $\eta_3 \rightarrow 1$. On the other hand, 
if $x_2 \rightarrow 0$ (with high probability), then $\eta_3 \rightarrow x_2$. 
Thus, we can identify (and partition) eight cases as follows: 
\begin{eqnarray}
{\rm Case} {\hspace{0.05in}} 1: & {\hspace{-0.05in}} 
x_2 \rightarrow 0, {\hspace{0.05in}} x_2 x_3 \rightarrow 0, 
{\hspace{0.05in}} x_2 x_3 x_4 \rightarrow 0 \Longrightarrow 
\eta_3 \rightarrow x_2, {\hspace{0.05in}} 
\eta_4 \rightarrow x_2 x_3 \Longrightarrow & 
{ \cal L} \rightarrow 0 \nonumber \\  
{\rm Case} {\hspace{0.05in}} 2: & {\hspace{-0.05in}} 
x_2 \rightarrow 0, {\hspace{0.05in}} x_2 x_3 \rightarrow 0, 
{\hspace{0.05in}} x_2 x_3 x_4 \rightarrow \infty \Longrightarrow 
\eta_3 \rightarrow x_2, {\hspace{0.05in}} 
\eta_4 \rightarrow x_2 x_3 \Longrightarrow & 
{ \cal L} \rightarrow \log(x_2 x_3 x_4) \nonumber \\  
{\rm Case} {\hspace{0.05in}} 3: & {\hspace{-0.05in}} 
x_2 \rightarrow 0, {\hspace{0.05in}} x_2 x_3 \rightarrow \infty, 
{\hspace{0.05in}} x_4 \rightarrow 0 \Longrightarrow 
\eta_3 \rightarrow x_2, {\hspace{0.05in}} 
\eta_4 \rightarrow 1 \Longrightarrow & 
{ \cal L} \rightarrow \log(x_2 x_3) \nonumber \\
{\rm Case} {\hspace{0.05in}} 4: & {\hspace{-0.05in}} 
x_2 \rightarrow 0, {\hspace{0.05in}} x_2 x_3 \rightarrow \infty, 
{\hspace{0.05in}} x_4 \rightarrow \infty \Longrightarrow 
\eta_3 \rightarrow x_2, {\hspace{0.05in}} 
\eta_4 \rightarrow 1 \Longrightarrow & 
{ \cal L} \rightarrow \log(x_2 x_3 x_4) \nonumber \\ 
{\rm Case} {\hspace{0.05in}} 5: & {\hspace{-0.05in}} 
x_2 \rightarrow \infty , {\hspace{0.05in}} x_3 \rightarrow 0, 
{\hspace{0.05in}} x_3 x_4 \rightarrow 0 \Longrightarrow 
\eta_3 \rightarrow 1, {\hspace{0.05in}} 
\eta_4 \rightarrow x_3 \Longrightarrow & 
{ \cal L} \rightarrow \log(x_2) \nonumber \\  
{\rm Case} {\hspace{0.05in}} 6: & {\hspace{-0.05in}} 
x_2 \rightarrow \infty , {\hspace{0.05in}} x_3 \rightarrow 0, 
{\hspace{0.05in}} x_3 x_4 \rightarrow \infty \Longrightarrow 
\eta_3 \rightarrow 1, {\hspace{0.05in}} 
\eta_4 \rightarrow x_3 \Longrightarrow & 
{ \cal L} \rightarrow \log(x_2 x_3 x_4) \nonumber \\  
{\rm Case} {\hspace{0.05in}} 7: & {\hspace{-0.05in}} 
x_2 \rightarrow \infty , {\hspace{0.05in}} x_3 \rightarrow \infty, 
{\hspace{0.05in}} x_4 \rightarrow 0 \Longrightarrow 
\eta_3 \rightarrow 1, {\hspace{0.05in}} 
\eta_4 \rightarrow 1 \Longrightarrow & 
{ \cal L} \rightarrow \log(x_2 x_3) \nonumber \\  
{\rm Case} {\hspace{0.05in}} 8: & {\hspace{-0.05in}} 
x_2 \rightarrow \infty , {\hspace{0.05in}} x_3 \rightarrow \infty, 
{\hspace{0.05in}} x_4 \rightarrow \infty \Longrightarrow 
\eta_3 \rightarrow 1, {\hspace{0.05in}} 
\eta_4 \rightarrow 1 \Longrightarrow & 
{ \cal L} \rightarrow \log(x_2 x_3 x_4) \nonumber 
\end{eqnarray}
In all the eight cases, we have a universal description 
for ${\cal L}$ (as $k \rightarrow \infty$) that holds with high 
probability: 
\begin{eqnarray}
{\cal L} \stackrel{ k \rightarrow \infty}{\approx} 
\sum_{m = 2}^{ \ell^{\star} - 1 } 
\log \big(x_m \big), {\hspace{0.1in}} 
\ell^{\star} = \arg \min_{2 \hspp \leq \hspp \ell\hspp \leq \hspp 4} 
\left\{ \prod_{m = \ell }^{ j } x_m \rightarrow 0 {\hspace{0.05in}} 
{\rm for} {\hspace{0.05in}} {\rm all} {\hspace{0.05in}} j \geq \ell 
\right\}. \nonumber 
\end{eqnarray}
If $\ell^{\star} = 2$, then the above summation is replaced by $0$, and if 
there exists no $\ell \in \{2,3,4 \}$ such that the above condition holds, 
then $\ell^{\star}$ is set to $5$.

The following proposition provides a precise mathematical formulation 
of the above heuristic. 
\begin{prop} 
\label{thm_mean_taua} 
Let the following limit be well-defined and be denoted as $\gamma_{\ell,j}$: 
\begin{eqnarray}
\gamma_{\ell, j} \triangleq \lim  \limits_{ k \rightarrow \infty}  
\frac{1}{k} \sum_{m = 1}^k \log \left( \frac{1 + \zeta_{m,j+1 } } 
{1 + \zeta_{m, \ell} }   \right). 
\nonumber 
\end{eqnarray}
Define $\ell^{\star}$ as 
\begin{eqnarray}
\label{ellstar}
\ell^{\star} & \triangleq & 
\arg \min_{\ell {\hspace{0.03in}} : 
{\hspace{0.03in}} 2 {\hspace{0.03in}} 
\leq {\hspace{0.03in}} \ell {\hspace{0.03in}} 
\leq {\hspace{0.03in}} L} \Big\{  
\Delta _{\ell,j} \leq 0 {\hspace{0.05in}} {\rm for} {\hspace{0.05in}} {\rm all} 
{\hspace{0.05in}} j = \ell, \cdots, L 
\Big\} 
{\hspace{0.05in}} {\rm where} \\ 
\Delta_{\ell, j} & = & \log \left( \frac{1 - \rho_{j, j+1} }
{ 1 - \rho_{\ell-1, \ell} } \right) + 
(j - \ell +1) D(f_1, f_0)
+  \gamma_{\ell,j}. 
\nonumber 
\end{eqnarray}
If there exists no element in the set for the $\arg\min$ operation 
in~(\ref{ellstar}), we set $\ell^{\star} = L+1$. 
Then, as $A \rightarrow \infty$ (and hence, 
$k = \nu_{ A } \rightarrow \infty$ a.s.\ from Prop.~\ref{prop_infty_bd}), 
we have 
\begin{eqnarray}
\frac{1}{k}  
\sum_{\ell = 2}^L \log \left( 1 + \eta_{\ell} x_{\ell}  \right) 
- \frac{1}{k} 
\sum_{\ell = 2}^{ \ell^{\star}-1 } \log( x_{\ell} ) 
\rightarrow 0 {\hspace{0.08in}} {\rm a.s. } 
\label{main_term}
\end{eqnarray}
If $\ell^{\star} = 2$, then the second term in the above expression is 
set to $0$. 
\end{prop}
\begin{proof} 
See Appendix~\ref{app_completionx}. 
\end{proof}

Following Props.~\ref{prop_infty_bd} and~\ref{thm_mean_taua}, 
as $A \rightarrow \infty$, $\nu_{ A }$ can be restated as 
\begin{eqnarray}
\nu_{ A } 
& \rightarrow &
\inf_k \Bigg \{ \sum_{m = 1}^k 
\Bigg( 
\log \left( \frac{1 - \rho_{1,2}}{1 - \rho} \right) + \log( L_{m,1} ) 
+ \log(1 + \zeta_{m,2}) + \frac{1}{k} \sum_{\ell = 2}^{\ell^{\star} - 1} 
\log( x_{\ell} ) \Bigg) 
> A \Bigg\} 
\nonumber \\ 
& = & \inf_k \Bigg \{ \sum_{m=1}^k \underbrace{  
\log \left( \frac{ 1 - \rho_{\ell^{\star} - 1, \ell^{\star} } } 
{1 - \rho} \right) + \sum_{ j = 1}^{\ell^{\star}-1 } \log( L_{m,j} ) 
+ \log \left( 1 + \zeta_{m, \ell^{\star} }  \right) 
}_{ y_m } > A \Bigg\} 
\label{eqn5} 
\end{eqnarray} 
with $\ell^{\star}$ defined in~(\ref{ellstar}). 

Observe that if the condition in Prop.~\ref{thm_mean_taua} is satisfied, 
the first $\ell^{\star}-1$ sensors contribute to the slope of $E_{\sf DD}$ 
and the rest of the sensors $\ell^{\star} , \cdots, L$ (if any) do 
not contribute to the slope. It is useful to understand the conditions 
under which 
$\ell^{\star} = L+1$. 

\ignore{ 
\begin{prop}
\label{prop_mu} 
For all $\ell = 1, \cdots, L - 1$, if there exists some $j \geq \ell+1$ 
such that 
\begin{eqnarray}
\label{cond_prop8}
(j - \ell) D(f_1, f_0) \geq \log \left( 
\frac{ \sum_{p = 0}^{\ell} (1 - \rho_{p, p+1})  } 
{1 - \rho_{j, j+1}} \right), 
\end{eqnarray}
we can bound $E[ \tau_{ {\sf NCTT}, A } ]$ as $A \rightarrow \infty$ as 
\begin{eqnarray}
\frac{E[  \tau_{ {\sf NCTT},A  } ]}{A} \leq \frac{1} 
{L D(f_1, f_0) + \log \left( \frac{1}{1 -\rho} \right) } . 
\nonumber 
\end{eqnarray} 
%
\end{prop}
}


Theorem~\ref{thm_final} provides a simple condition such that the 
observations from all the $L$ sensors contribute to the slope. We are 
now prepared to prove it. 

\noindent {\bf \em Proof of Theorem~\ref{thm_final} ($L \geq 3$):} First, 
using Lemma~\ref{lem_bd} note that, we can bound $\Delta_{\ell,j}$ as 
\begin{eqnarray}
\Delta_{\ell,j } \geq 
(j - \ell + 1) D(f_1, f_0) + \log(1 - \rho_{j, j+1}) 
- E \left[ \log \left( \sum_{p = 0}^{\ell - 1} 
\frac{ (1 - \rho_{p, p +1}) }
{ \prod_{i = p+1}^{\ell - 1} L_{\bullet, i} } \right) \right]. 
\nonumber 
\end{eqnarray}
Using Jensen's inequality and noting that $E_{f_1} \left[ \frac{1}
{ \prod_{i = p+1}^{\ell-1} L_{\bullet, i} } \right] = 1$,~(\ref{cond_five}) 
is sufficient to ensure that for all $\ell = 2, \cdots, L$, there 
exists some $j \geq \ell$ such that $\Delta_{\ell,j} > 0$. It is 
important to realize that the above condition is necessary as well as 
sufficient for $\ell^{\star} = L + 1$. 
Thus, under the assumption that~(\ref{cond_five}) holds, 
invoking Prop.~\ref{prop_infty_bd} as $A \rightarrow \infty$ 
(that is, letting $k = \nu_{ A} \rightarrow \infty$ a.s.\ 
and using Prop.~\ref{thm_mean_taua}), $\nu_{ A }$ can be written as 
\begin{eqnarray}
\nu_{ A } \stackrel{A \rightarrow \infty} {\rightarrow} 
\inf_k \left\{ \sum_{m=1}^k \left( 
\sum_{\ell=1}^L \log(L_{m,\ell} ) + 
\log \left( \frac{1}{1 - \rho} \right) + \log( 1 + \zeta_{m, L+1} )
\right) > A \right\}. \nonumber
\end{eqnarray}
Note that since $\zeta_{m,L+1} \geq 0$, we have 
\begin{eqnarray}
\sum_{m=1}^k \left( \sum_{\ell=1}^L \log(L_{m,\ell}) + 
\log \left( \frac{1}{1 - \rho} \right) + \log( 1 + \zeta_{m,L+1} ) 
\right) \geq 
\underbrace{
\sum_{m=1}^k \left( \sum_{\ell=1}^L \log(L_{m,\ell} ) + 
\log \left( \frac{1}{1 - \rho} \right) \right) 
}_{ L_k  } , \nonumber 
\end{eqnarray}
and hence, $\nu_{ A } \leq \nu_{L,A}$ where 
\begin{eqnarray}
\nu_{L,A} \triangleq  \inf_k \Big \{ L_k >A \Big \}. \nonumber
\end{eqnarray}
Thus, we have 
\begin{eqnarray}
\frac{E[ \nu_{ A } ]} {A} \leq 
\frac{E[ \nu_{L,A}] } {A} \stackrel{A \rightarrow \infty}{\rightarrow} 
\frac{1}{ L D(f_1, f_0) + \log \left( \frac{1}{1-\rho}\right) } 
\nonumber 
\end{eqnarray}
where the convergence is again due to Lemma~\ref{lem_blackwell}. 
\endproof

\section{Discussion and Numerical Results} 
\label{sec8} 
\noindent {\bf \em Discussion:} A loose sufficient condition for all the 
$L$ sensors to contribute to the slope of $E_{\sf DD}$ of $\nu_A$ is that 
\begin{eqnarray}
D(f_1, f_0) > \max \limits_{\ell = 1, \cdots, L-1} 
{\hspace{0.04in}} \min \limits_{j \geq \ell+1} 
{\hspace{0.04in}} \frac{1}{j - \ell} 
\cdot \log \left(  \frac{ \sum_{p = 0}^{\ell} (1 - \rho_{p, p+1} ) } 
{1 - \rho_{j, j+1}  } \right) \triangleq \gamma_u.  
\nonumber 
\end{eqnarray}
Another sufficient condition is that 
\begin{eqnarray}
D(f_1, f_0) > \max \limits_{ \ell = 1, \cdots, L-1} 
\frac{1}{L - \ell} \cdot \log \left( 1  - \rho + \sum_{j = 1}^{\ell} 
(1 - \rho_{j, j+1})  \right). \nonumber 
\end{eqnarray}
That is, if $\rho$ is such that 
\begin{eqnarray}
\rho \geq \sum_{\ell = 2}^L (1 - \rho_{\ell-1, \ell}), \nonumber 
\end{eqnarray}
then $\gamma_u \leq 0$ and the condition of Theorem~\ref{thm_final} 
reduces to a mild one that the K-L divergence between $f_1$ 
and $f_0$ be positive. A special setting where the above condition is true 
(irrespective of the rarity of the disruption-point) is the regime where 
change propagates 
across the sensor array ``quickly.'' The case of~\cite{venu_qdecentral} is an 
extreme example of this regime and Theorem~\ref{thm_final} recaptures this 
extreme case. 

In more general regimes where change propagates across the sensor array 
``slowly'', either the disruption-point should become less rare 
(independent of the choice of $f_1$ and $f_0$) or that the densities 
$f_1$ and $f_0$ be sufficiently discernible (independent of the rarity 
of the disruption-point) so that all the $L$ sensors can contribute to 
the asymptotic slope. When these conditions fail to hold, it is not 
clear whether the theorems are applicable, or even if all the $L$ 
sensors contribute to the slope of $E[\nu_{  A}]$. Nevertheless, it is 
reasonable to conjecture that as long as $\min \limits_{\ell} 
\rho_{\ell-1,\ell} > 0$, then all the $L$ sensors contribute to the 
asymptotic slope. 

However, the difference between the asymptotic and the non-asymptotic 
regimes need a careful revisit. Following the initial remark 
(Prop.~\ref{prop_blocking}) on the extreme case of blocking sensors 
(where some $\rho_{\ell-1, \ell} = 0$), in the more realistic case 
where some $\rho_{\ell-1, \ell}$ may be small (but non-zero), it is 
possible that if $D(f_1, f_0)$ is smaller than some threshold value 
(determined by the change propagation parameters), not all of the $L$ 
sensors may ``effectively'' contribute to the slope of $E_{\sf DD}$, 
at least for reasonably small, but non-asymptotic values of $P_{\sf FA}$. 
For example, see the ensuing discussion where numerical results 
illustrate this behavior at $P_{\sf FA}$ values of $10^{-4}$ to $10^{-5}$ 
for some choice of change propagation parameters, {\em even} when the 
condition in Theorem~\ref{thm_final} is met. When the condition in 
Theorem~\ref{thm_final} is not met, such a behavior is expected to be more 
typical. 

The final comment is on the approach pursued in this paper. While the 
approach pursued in Sec.~\ref{sectionxx} and~\ref{sectionxxx} results in 
interesting conclusions, it is not clear if this approach is {\em fundamental} 
in the sense that this is the only approach possible for characterizing 
$E_{\sf DD}$ vs.\ $P_{\sf FA}$. 
Furthermore, this approach assumes the existence of $\{ \gamma_{\ell,j } \}$. 
Even if these quantities exist and are hence, theoretically computable, such a 
computation is complicated by the fact that $\{ \zeta_{m,\ell}, 
{\hspace{0.05in}} m = 1, \cdots, k \}$ are correlated. Thus, verification 
of the exact condition in Prop.~\ref{thm_mean_taua} (equivalently, computing 
$\ell^{\star}$) has to be achieved either via Monte Carlo methods or by 
bounding $\Delta_{\ell,j}$, as done here. Furthermore, correlation of $\{ \zeta_{m, \ell} \}$ 
and hence, $y_m$ (see~(\ref{eqn5})) implies that statistics of $\nu_{ A }$ have 
to be obtained using non-linear renewal theoretic techniques for general 
(correlated) random variables~\cite{woodroofe}. This is the subject of 
current work.

\ignore{ 
[More discussion... as $\min_{\ell \geq 1} \rho_{\ell-1,\ell}$ increases 
towards $1$, $\ell^{\star}$ moves towards $L$ irrespective of 
$D(f_1, f_0)$... For any choice of $\min_{\ell \geq 1} \rho_{\ell-1,\ell} 
< 1$, there exists a cut-off divergence below which not all sensors 
contribute to slope. Need to present at least some numerical results to 
clearly show this behavior. Working on this.] 
}

\noindent {\bf \em Numerical Study I -- Performance Improvement with 
$\nu_{A }$:} Given that the structure of $\tau_{\sf opt}$ is not known in 
closed-form, we now present numerical studies to show that $\nu_A$ results 
in substantial improvement in performance over both a single sensor test 
(which uses the observations only from the first sensor and ignores the 
other sensor observations) and a test that uses the observations from all 
the sensors but under a mismatched model (where the change-point for all 
the sensors is assumed to be the same), even under realistic modeling 
assumptions. 

\begin{figure}[htb!]
\centering
\begin{tabular}{c}
\includegraphics[height=3.5in,width=3.8in]{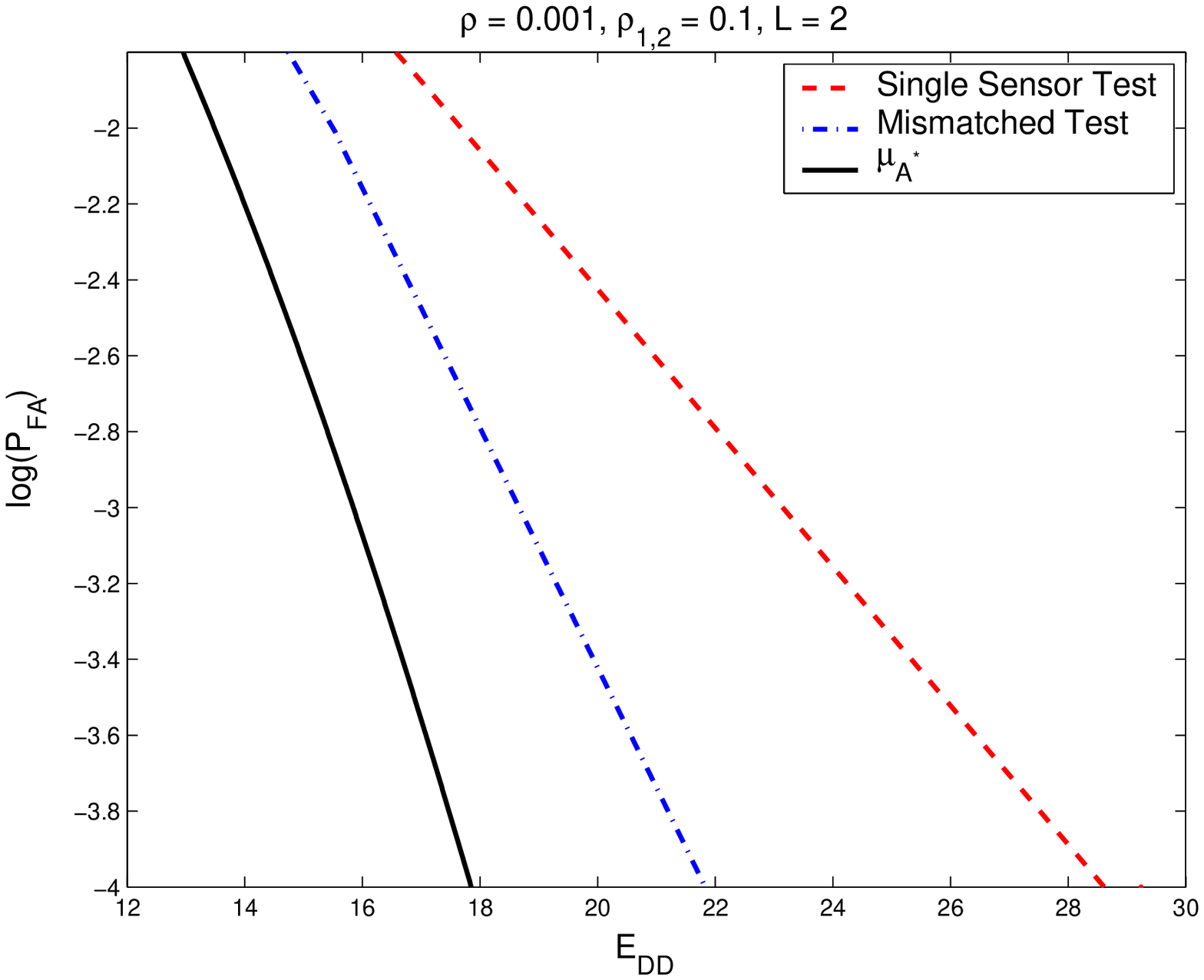}
\end{tabular}
\caption{False alarm vs.\ Expected detection delay for a $L = 2$ setting 
with $\rho = 0.001$ and $\rho_{1,2} = 0.1$. \label{fig1x}}
\end{figure}

The first example corresponds to a two sensor system where the 
occurrence of change is modeled as a geometric random variable with 
parameter $\rho = 0.001$. Change propagates from the first sensor to 
the second with the geometric parameter $\rho_{1,2} = 0.1$. The 
pre- and post-change densities are ${\cal CN}(0,1)$ and ${\cal CN}(1,1)$, 
respectively so that $D(f_1, f_0) = 0.50$. Fig.~\ref{fig1x} shows that 
$\nu_{A}$ can result in an improvement of at least $4$ units of 
delay at even marginally large $P_{\sf FA}$ values on the order of $10^{-3}$.

\begin{figure}[htb!]
\centering
\begin{tabular}{c}
\includegraphics[height=3.5in,width=3.8in]{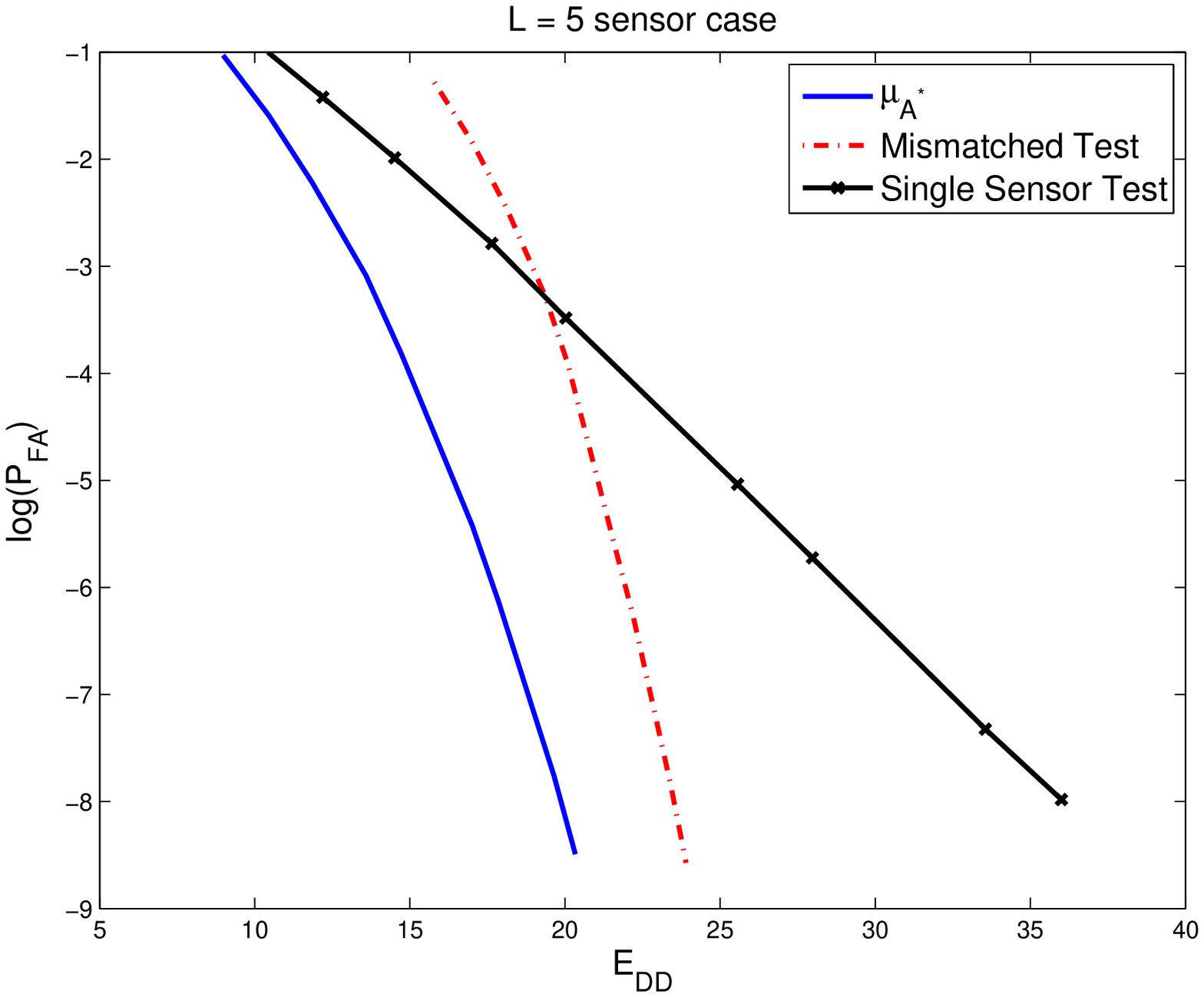}
\end{tabular}
\caption{False alarm vs.\ Expected detection delay for a typical 
$L = 5$ setting. 
\label{fig_fivesensor}}
\end{figure}

The second example corresponds to a five sensor system where $\rho = 0.005$. 
Change propagates across the array according to the following model: 
$\rho_{1,2} = 0.1, \rho_{2,3} = 0.2, \rho_{3,4} = 0.5$ and $\rho_{4,5} = 0.7$. 
The pre- and the post-change densities are ${\cal CN}(0,1)$ and ${\cal CN}(0.75,1)$ 
so that $D(f_1, f_0) \approx 0.2813$. With $D(f_1, f_0)$ and the change 
parameters as above, Theorem~\ref{thm_final} assures us that at least $L = 2$ 
sensors contribute to the $E_{\sf DD}$ vs.\ $P_{\sf FA}$ slope asymptotically. 
On the other hand, Fig.~\ref{fig_fivesensor} shows that more than two sensors 
indeed contribute to the slope. Thus, it can be seen that 
Theorem~\ref{thm_final} provides only a sufficient condition on performance 
bounds. It is also worth noting the transition in slope (unlike the case 
in~\cite{venu_qdecentral}) for both the mismatched 
test and $\nu_{A}$ as $P_{\sf FA}$ decreases from moderately large 
values to zero, whereas the slope of the single sensor test (as expected) remains 
constant.

\noindent {\bf \em Numerical Study II -- Performance Gap Between the 
Tests:} We now present a 
second case-study with the main goal being the understanding of the 
relative performance of $\nu_{A}$ with respect to the single sensor and 
the mismatched tests. We again consider a $L = 2$ sensor system and we 
vary the change process parameters, $\rho$ and $\rho_{1,2}$, in this 
study. The pre- and the post-change densities are ${\cal CN}(0,1)$ and 
${\cal CN}(1.2,1)$ so that $D(f_1, f_0) = 0.72$. 

\begin{figure}[htb!]
\begin{center}
\begin{tabular}{cc}
\includegraphics[height=2.5in,width=3in]{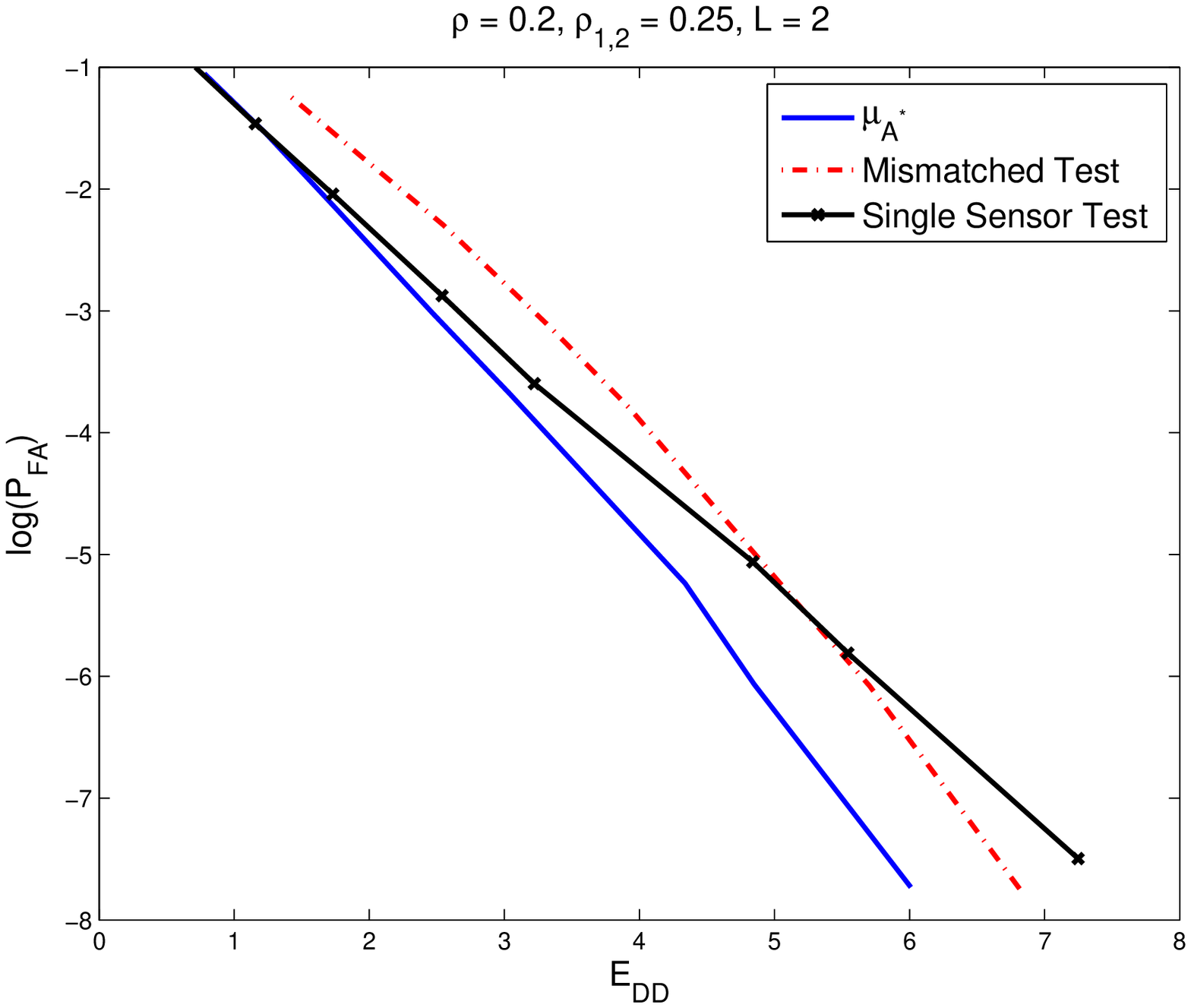} & 
\includegraphics[height=2.5in,width=3in]{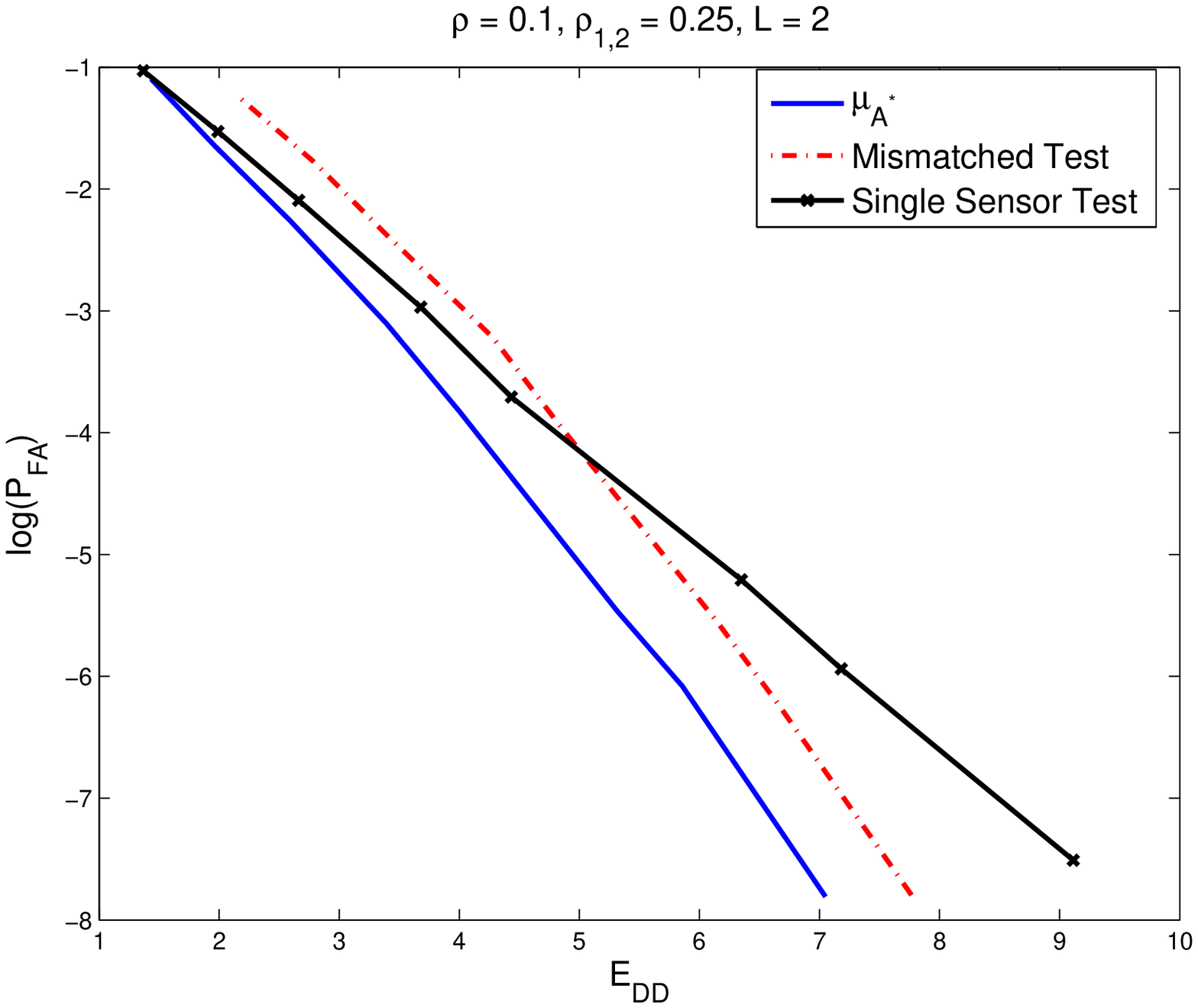}
\\ (a) & (b) \\ 
\includegraphics[height=2.5in,width=3in]{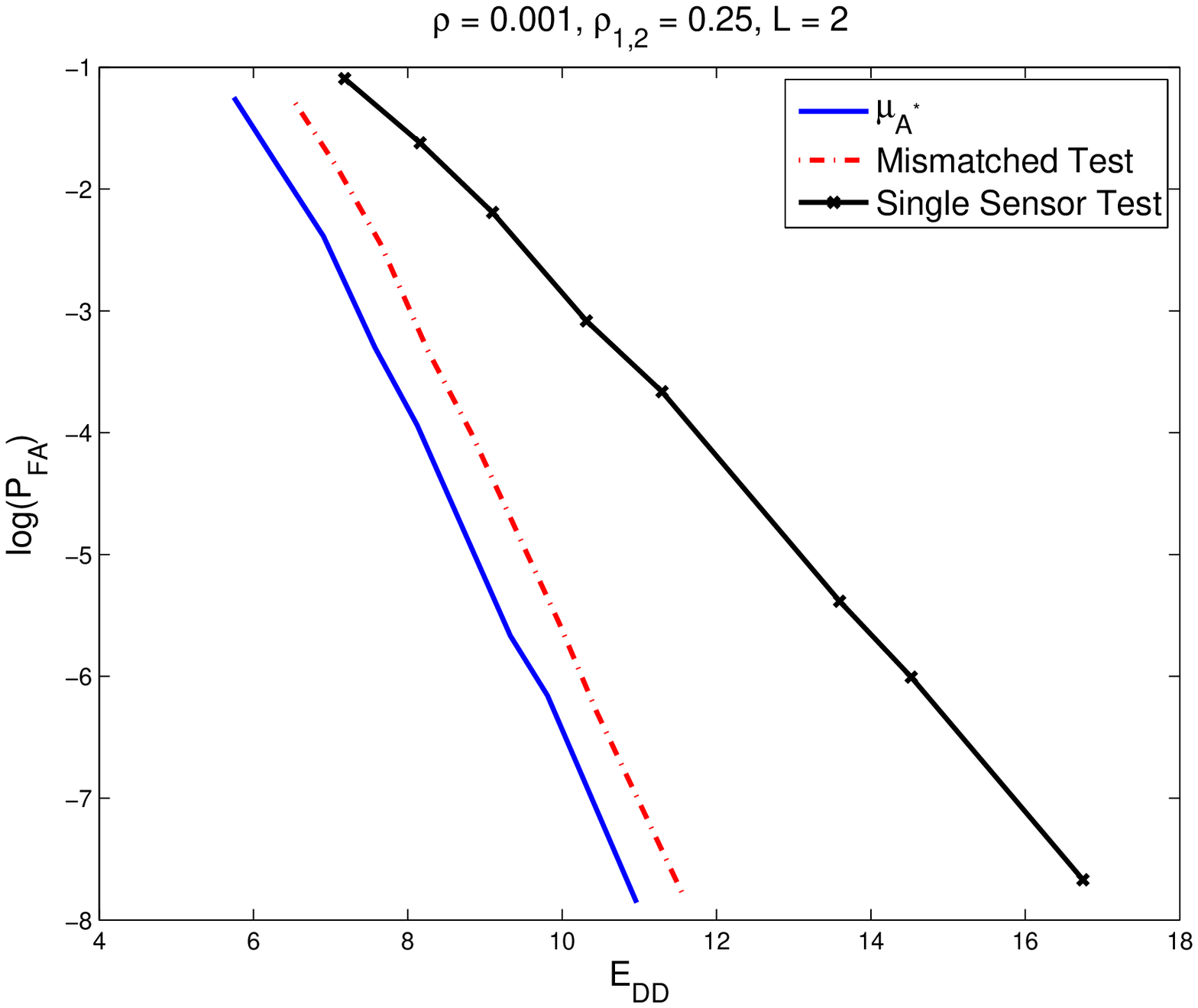} & 
\includegraphics[height=2.5in,width=3in]{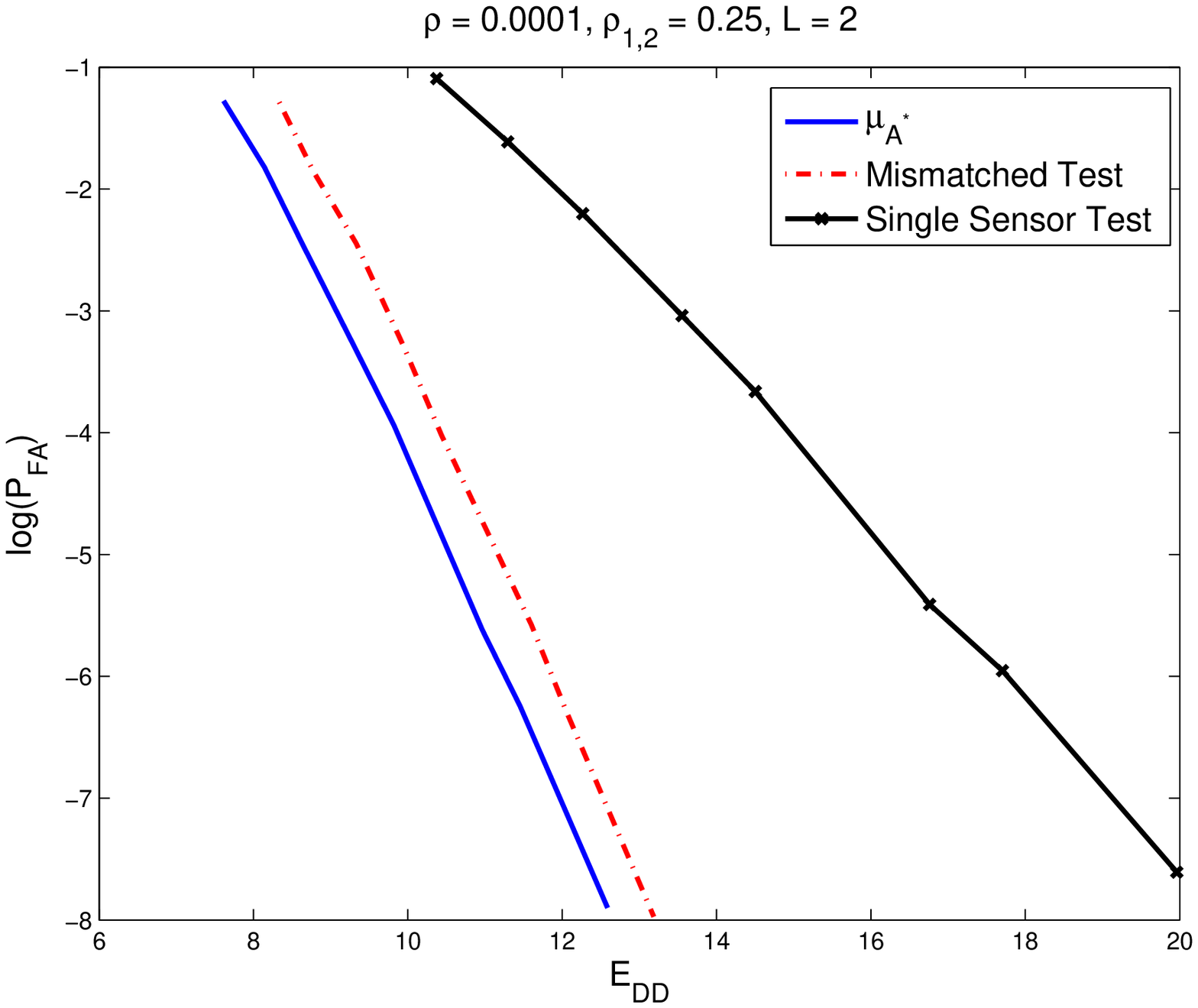}
\\ (c) & (d) 
\end{tabular} 
\caption{\label{fig1} False alarm vs.\ Expected detection delay for 
a $L = 2$ setting with different model parameters.} 
\end{center}
\end{figure}

\begin{figure}[htb!]
\begin{center}
\begin{tabular}{cc}
\includegraphics[height=2.5in,width=3in]{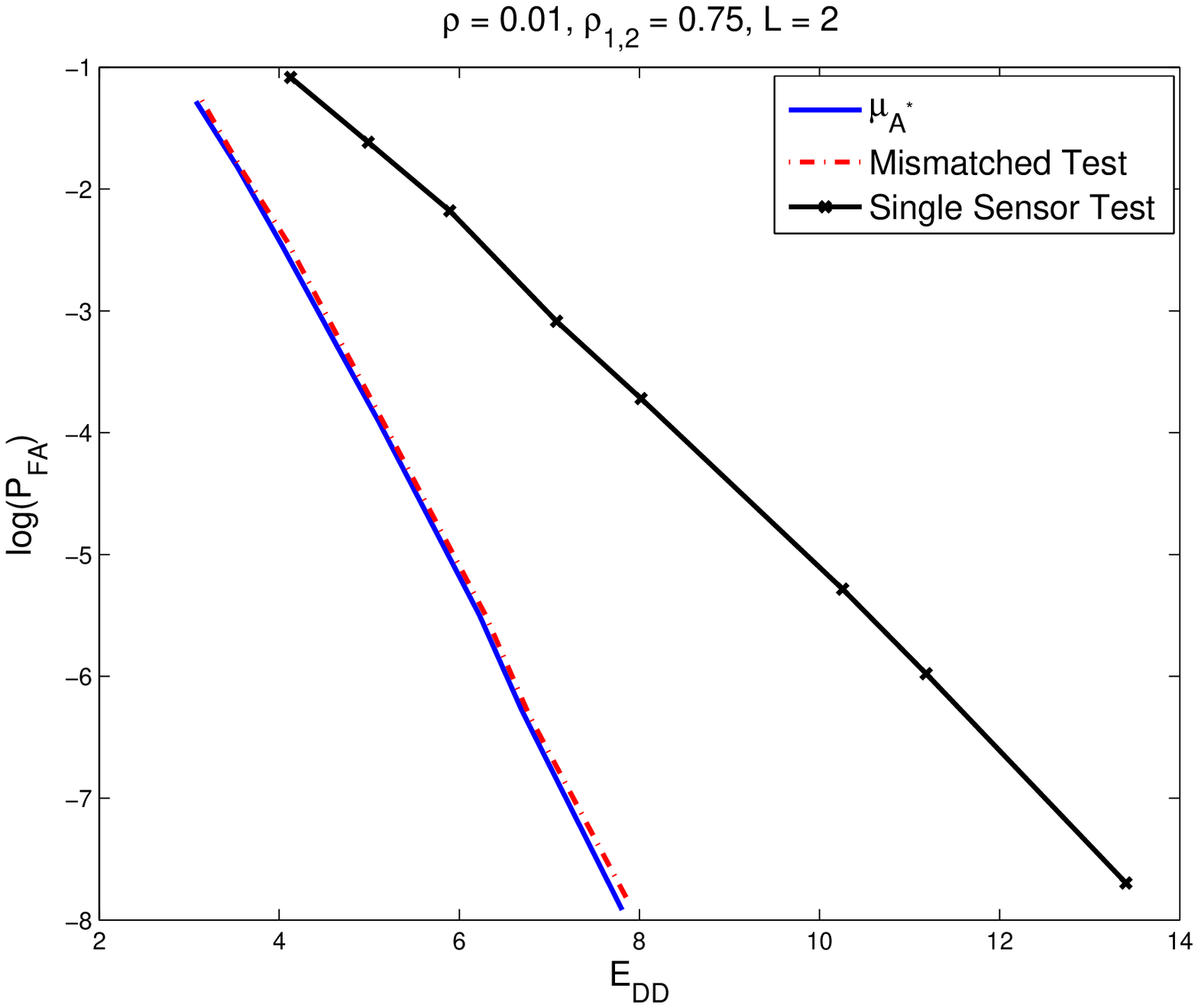} & 
\includegraphics[height=2.5in,width=3in]{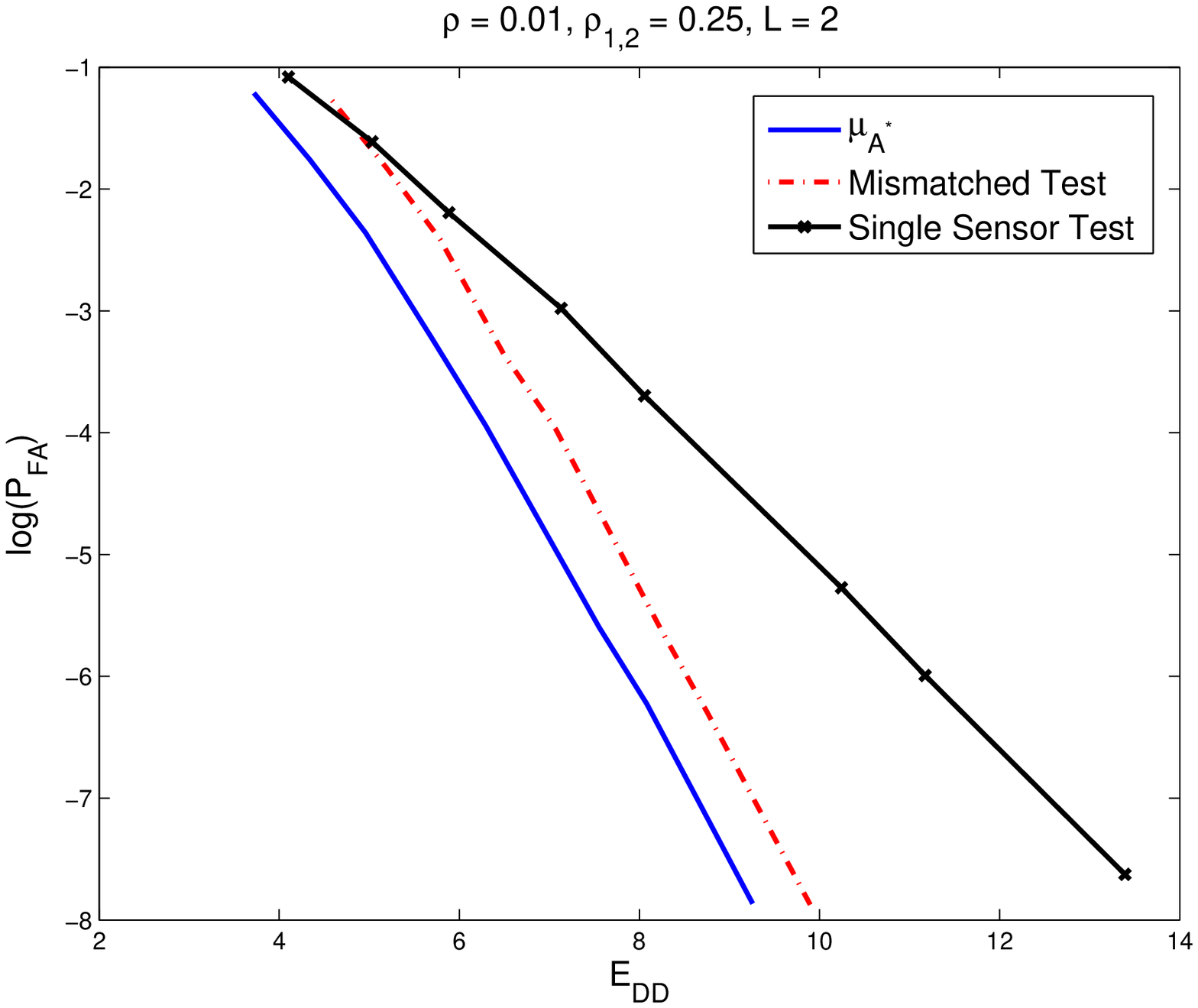}
\\ (a) & (b) \\ 
\includegraphics[height=2.5in,width=3in]{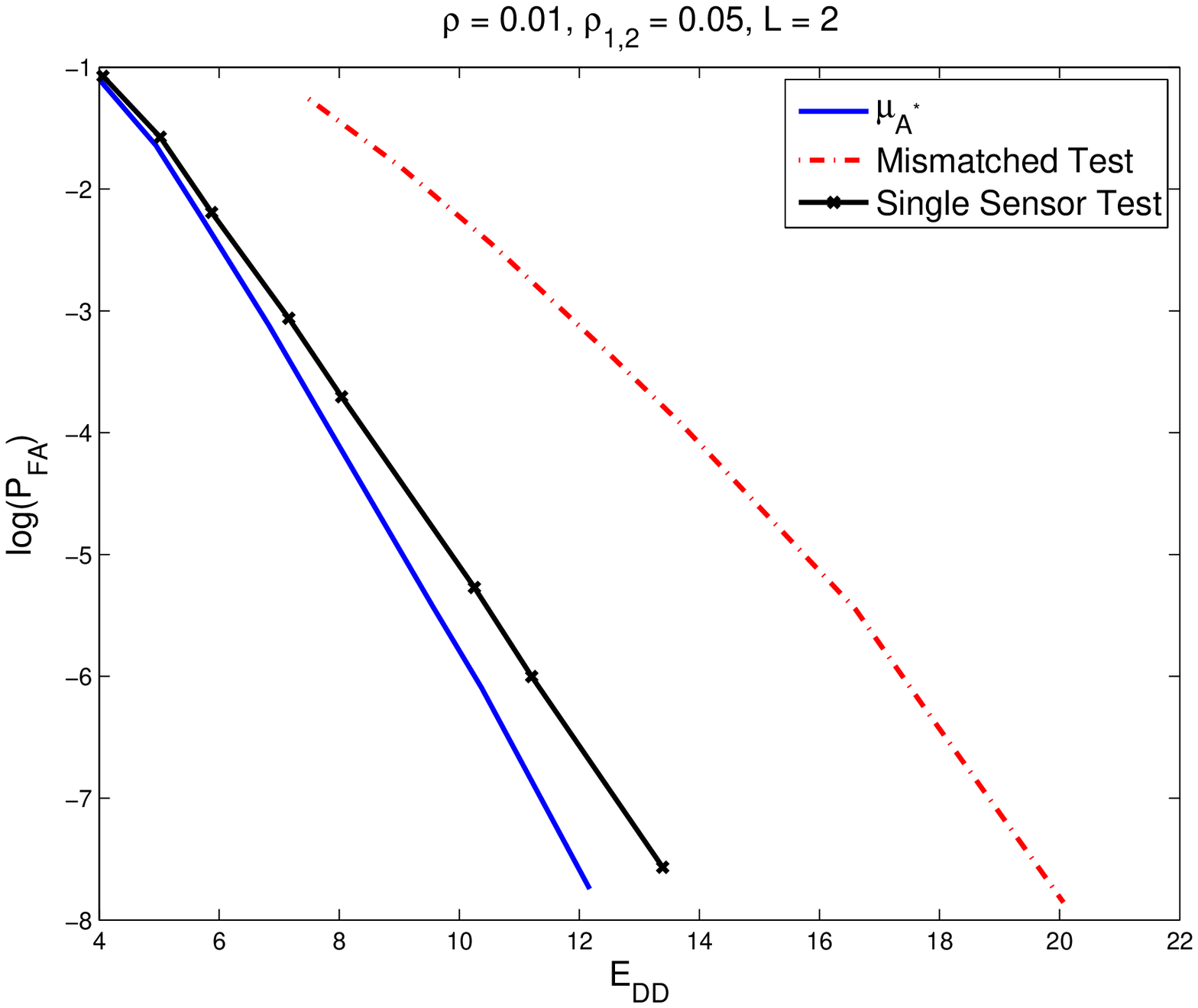} & 
\includegraphics[height=2.5in,width=3in]{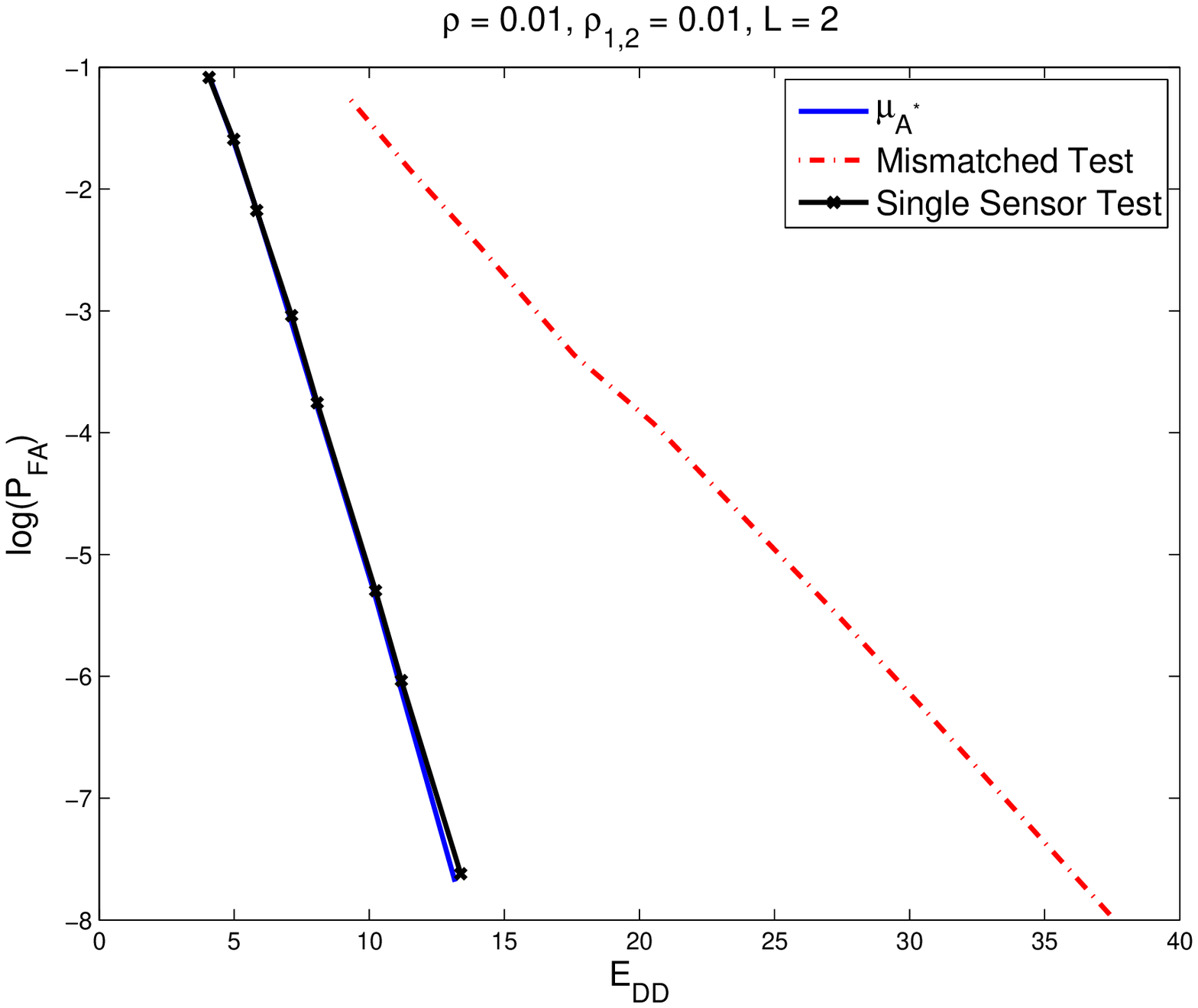}
\\ (c) & (d) 
\end{tabular} 
\caption{\label{fig2} False alarm vs.\ Expected detection delay for 
a $L = 2$ setting with different model parameters.} 
\end{center}
\end{figure}

Fig.~\ref{fig1} and Fig.~\ref{fig2}(b) show the performance of the 
three tests with varying $\rho$ parameters for a fixed choice of 
$\rho_{1,2}$. We observe that the gap in performance between the 
single sensor test and $\nu_{A}$ 
increases as $\rho$ decreases, whereas 
the gap between $\nu_{A}$ and the mismatched test stays fairly constant. 
Similarly, Fig.~\ref{fig2} shows the performance of the three tests 
with varying $\rho_{1,2}$ parameters for a fixed choice of $\rho$. 
We observe from these plots that the gap between the mismatched test 
and $\nu_{A}$ increases as $\rho_{1,2}$ decreases, whereas the gap between 
the single sensor test and $\nu_{A}$ increases as $\rho_{1,2}$ increases.

The choice of $D(f_1, f_0) = 0.72$ is such that the sufficient 
condition in Theorem~\ref{thm_final} are satisfied, independent of the 
change parameters. Hence, we expect 
the slope of the $E_{\sf DD}$ vs.\ $P_{\sf FA}$ plot to be of the form 
$\frac{1}{2D(f_1, f_0)  + |\log(1 - \rho)| }$ 
{\em asymptotically} as $P_{\sf FA} \rightarrow 0$. Nevertheless, 
Fig.~\ref{fig2}(c) and (d) show that, when both $\rho$ and $\rho_{1,2}$ 
are small, the slope of $\nu_{A}$ is only as good as (or slightly 
better than) the single sensor test, which is known to have a slope of 
the form $\frac{1}{D(f_1, f_0)  + |\log(1 - \rho)| }$. Thus, we see that 
even though our theory guarantees that both the sensors' observations 
contribute in the eventual performance of $\nu_{A}$ 
{\em asymptotically}, 
we may not see this behavior for reasonable choices of $P_{\sf FA}$ like 
$10^{-4}$. The case of observation models not meeting the conditions of 
Theorem~\ref{thm_final} is expected to show this trend for even lower 
$P_{\sf FA}$ values. 

To summarize these observations, if $E_{\sf DD, {\hspace{0.03in}} \nu_{A} 
}$, $E_{\sf DD, {\hspace{0.03in}} MM }$ and 
$E_{\sf DD, {\hspace{0.03in}}  SS}$ denote the expected detection delays 
for $\nu_{A}$, mismatched and single sensor tests (respectively) 
for some fixed choice of $P_{\sf FA}$, then 
\begin{eqnarray}
E_{\sf DD, {\hspace{0.03in}}  MM } - 
E_{\sf DD, {\hspace{0.03in}}  \nu_{A} } 
& \propto & \frac{1}{\rho_{1,2}} 
{\hspace{0.05in}} {\rm and} {\hspace{0.05in}} 
{\rm independent} {\hspace{0.05in}} {\rm of} {\hspace{0.05in}} \rho
\nonumber \\ 
E_{\sf DD, {\hspace{0.03in}}  SS } - 
E_{\sf DD, {\hspace{0.03in}}  \nu_{A} } & \propto & 
\frac{\rho_{1,2}}{\rho}. \nonumber 
\end{eqnarray}

It is interesting to note from the above equations that $\rho_{1,2}$ 
impacts the gap between the two tests in a contrasting way. The test 
$\nu_{A}$ is 
expected to result in significant performance improvement in the regime 
where $\rho$ is small, but $\rho_{1,2}$ is neither too small nor too 
large. In fact, this regime where $\nu_{A}$ 
is expected to result in significant 
performance improvement is the precise regime that is of importance in 
practical contexts. This is so because we can expect the occurrence of 
disruption (e.g., cracks in bridges, intrusions in networks, onset of 
epidemics etc.) to be a rare phenomenon. Once the disruption occurs, we 
expect change to propagate across the sensor array fairly quickly due to 
the geographical (network proximity in the case of computer networks) 
proximity of the other sensors, but not so quick that the extreme case 
of~\cite{venu_qdecentral} is applicable. Classifying the regime of 
$\{\rho_{\ell-1,\ell}  \}$ and $D(f_1, f_0)$ where significant 
performance improvement is possible with $\nu_{A}$ is ongoing work. It 
is also of interest to come up with better test structures in the regime 
where $\nu_{A}$ does not lead to a significant performance improvement.

\section{Concluding Remarks}
\label{sec9} 
We considered the centralized, Bayesian version of the change process 
detection problem in this work and posed it in the classical POMDP 
framework. This formulation of the change detection problem allows 
us to establish the sufficient statistics for the DP under study and a 
recursion for the sufficient statistics. While we obtain the broad 
structure of the optimal stopping rule ($\tau_{\sf opt}$), any further 
insights into it are rendered infeasible by the complicated nature of 
the infinite-horizon cost-to-go function. Nevertheless, $\tau_{\sf opt}$ 
reduces to a threshold rule (denoted in this work as $\nu_{A}$) in the 
rare disruption regime. The test $\nu_{A}$ possesses many attractive 
properties: i) it is of low-complexity; ii) it is asymptotically optimal 
in the vanishing false alarm probability regime under certain mild 
assumptions on the K-L divergence between the post- and the pre-change 
densities; and iii) numerical studies suggest that it can lead to 
substantially improved performance 
over naive tests. Thus, $\nu_{A}$ serves as an attractive test for practical 
applications that can be modeled as a change process.

To the best of our knowledge, this is the first work to consider the 
change process detection problem in extensive detail. Thus, there 
exists potential for extending this work in multiple new directions. 
While we established the asymptotic optimality of $\nu_{A}$ when 
$D(f_1, f_0) \geq \gamma_u$, it is unclear as to what happens when 
$D(f_1, f_0) < \gamma_u$. In other words, is $\ell^{\star} = L + 1$ when 
$D(f_1, f_0) < \gamma_u$ given that $\gamma_u > 0$? It is most likely 
that $\nu_{A}$ is asymptotically optimal even in 
this regime as long as $\min \limits_{\ell} \rho_{\ell - 1, \ell} > 0$, 
but establishing this result may involve some ingenious techniques. 
However, if $\nu_{A}$ is not asymptotically optimal in this regime, it 
is of interest to design better low-complexity stopping rules; e.g., 
Threshold tests on weighted sums of the {\em a posteriori} probabilities 
based on further study of the structure of $\tau_{\sf opt}$ etc. 

More careful asymptotic analysis of $\nu_{A}$ and performance gap 
between: i) $\nu_{A}$ and the mismatched test, ii) $\nu_{A}$ and the 
single sensor test, and iii) $\nu_{A}$ and weighted threshold tests 
etc.\ would involve tools from non-linear renewal 
theory~\cite{woodroofe,TVAS,tart2} and is the subject of current 
attention. Such an asymptotic study could in turn drive the design 
of better test structures. Our numerical results also illustrate and 
motivate the need for non-asymptotic characterization (piece-wise 
linear approximations of the $E_{\sf DD}$ vs.\ $P_{\sf FA}$ curve) 
of the proposed tests. Unlike the case of instantaneous change 
propagation~\cite{venu_qdecentral,TVAS}, we showed that asymptotic 
characterizations may not kick in quickly for small $P_{\sf FA}$ 
values if the change propagates too ``slowly'' across the sensor 
array. Under such circumstances, it is also of interest to revisit 
the precise definition of optimality of a stopping rule. 

\ignore{ 
This work is {\em critically} dependent on the twin assumptions that: 
i) The observations can be obtained perfectly at the fusion center, and 
ii) The statistics of the change process can be learned {\em a priori} 
at the fusion center. While these assumptions are realistic in certain 
applications, it is important to address more general versions of 
the problem considered in this work. For example, energy, power and 
delay constraints between the sensors and the fusion center are 
possible in many applications. More general array geometries than a 
line of sensors and uncertainties in the location of the disruption-causing 
agent(s) and the disruption dynamics could result in uncertainties in 
learning the change process model. Despite the simplifications that a 
Markovian model offers, the change process setup is characterized by a large 
number of parameters which not only makes analysis cumbersome 
(Sec.~\ref{sectionxxx}), but also puts constraints on learning these 
parameters accurately. 
}

Decentralized~\cite{venu_sequential,venu_qdecentral}, censored~\cite{swaroop}, 
multi-channel~\cite{Taveer} and robust~\cite{venu_robust,huber} 
versions of change detection are motivated by these constraints. 
Extensions of this work to more general observation models are 
important in the context of practical applications. For example, 
non-iid~\cite{TVAS} and Hidden-Markov models~\cite{Cdfuh} have found 
increased interest in biological problems determined by an event-driven 
potential~\cite{farwell,ratnam}. Practical applications will in turn 
drive the need for understanding change detection with certain 
specific observation models. 


\appendix 
\subsection{Completing Proof of Theorem~\ref{thm_structure}: 
Establishing Concavity of 
$A_k^T(\cdot)$ and $J_k^T(\cdot)$} 
\label{app_ccv}
We now show that $A_k^T(\pb_k)$ and $J_k^T(\pb_k)$ are concave in $\pb_k$. First, 
note that $J_T^T(\pb_T) = p_{T,1}$ is concave in $\pb_T$ because it is affine. 
Using the recursion for $\pb_T$, it is straightforward to check that 
\begin{eqnarray}
A_{T-1}^T(\pb_{T-1}) = 
E[J_{T}^T(\pb_T)|I_{T-1}] & = & p_{T-1, 1} \cdot (1 - \rho).
\nonumber 
\end{eqnarray}
Using this in the definition of $J_{T-1}^T(\pb_{T-1})$, we have 
\begin{eqnarray}
J_{T-1}^T(\pb_{T-1}) 
& 
= & \left\{ \begin{array}{cc} 
p_{T-1,1} & 0 \leq p_{T-1,1} \leq \frac{c}{c + \rho} \\ 
c +  p_{T-1,1}(1 - \rho - c) & \frac{c}{c+\rho} \leq p_{T-1,1} \leq  1. 
\end{array}
\right. \nonumber 
\end{eqnarray} 
Since both $A_{T-1}^T(\pb_{T-1})$ and $J_{T-1}^T(\pb_{T-1})$ are affine 
and piecewise-affine (It is important to note that the slope of the 
second affine part, which is $1 - \rho -c$, is smaller than the 
first ($= 1$).) in $\pb_{T-1,1}$ respectively, they are concave.

We now assume that $J_{k+1}^T(\pb_{k+1})$ is concave in $\pb_{k+1}$ and 
show that $A_k^T(\pb_k)$ is also concave in $\pb_k$. For this, consider 
$\lambda 
A_k^T(\pb_{k}^1) + (1 - \lambda) A_k^T(\pb_{k}^2)$ with $\pb_{k}^1$ and 
$\pb_{k}^2$ being two elements in the standard $L$-dimensional 
simplex. We have 
\begin{eqnarray}
\lambda A_k^T(\pb_{k}^1) + (1 - \lambda) A_k^T(\pb_{k}^2)   
&=& \int 
\Big[ \lambda J_{k+1}^T \left( \pb_{k+1}^1 \right) 
\mu_1 
+ 
(1 -\lambda) J_{k+1}^T \left( \pb_{k+1}^2 \right) 
\mu_2 
\Big]  \Big|_{\Zb_{k+1} = \zb} d\zb \nonumber \\ 
&=& \int 
\Big[ \mu J_{k+1}^T \left( \pb_{k+1}^1 \right) + 
(1 - \mu) J_{k+1}^T \left( \pb_{k+1}^2 \right) \Big] 
\nonumber \\& & {\hspace{0.1in}} 
\times 
\Big( \lambda \mu_1 + (1 - \lambda) \mu_2 
\Big)
\Big|_{\Zb_{k+1} = \zb} d\zb \nonumber 
\end{eqnarray}
where 
\begin{eqnarray} 
\mu_i & = & f(\Zb_{k+1} | I_k )  \Big|_{ \pb_k = \pb_k^i } 
= \sum_{j=1}^{L+1} \left[ 
\left(  \sum_{m=1}^j w_{k+1,j,m} {\hspace{0.04in}} 
p_{k,m}^i 
\right) {\bf \Phi}_{\sf obs}(k+1,j) \right], 
{\hspace{0.05in}} i =1,2, {\hspace{0.04in}} {\rm and} 
\nonumber \\  
\mu & = & \frac{ \lambda \mu_1 } 
{ \lambda \mu_1 + (1 - \lambda) \mu_2 }. \nonumber 
\end{eqnarray}

Using the concavity of $J_{k+1}^T(\cdot)$, we can upper bound the 
above as follows: 
\begin{eqnarray}
\lambda A_k^T(\pb_{k}^1) + (1 - \lambda) A_k^T(\pb_{k}^2)   
& \leq & 
\int \Big[ J_{k+1}^T \Big( \mu \pb_{k+1}^1 + (1- \mu) \pb_{k+1}^2 \Big) 
\nonumber \\ 
&& {\hspace{0.15in}} \times \Big( 
\lambda \mu_1 + (1 - \lambda) \mu_2 \Big) \Big] \Big|_{\Zb_{k+1} = \zb} 
d\zb \nonumber 
\end{eqnarray}
If we define 
\begin{eqnarray}
\pb_k^3 \triangleq \lambda \pb_k^1 + (1 - \lambda) \pb_k^2, \nonumber 
\end{eqnarray}
it is straightforward to check that 
\begin{eqnarray}
\pb_{k+1}^3  & = &  \mu \pb_{k+1}^1 + (1 - \mu) \pb_{k+1}^2.  \nonumber 
\end{eqnarray} 
Using these facts, we have 
\begin{eqnarray}
\lambda A_k^T(\pb_{k}^1) + (1 - \lambda) A_k^T(\pb_{k}^2)   
\leq A_k^T(\lambda\pb_k^1 + (1 - \lambda) \pb_k^2), \nonumber 
\end{eqnarray} 
thus establishing the concavity of $A_k^T(\cdot)$. The concavity of $J_k^T(\cdot)$ 
follows since the minimum and sum of concave functions is concave. 
An inductive argument completes the proof. 
\endproof 

\subsection{Proof of Theorem~\ref{thm_rho0}}
\label{app_rho0}
We will show that 
\begin{eqnarray}
\tau_{\sf opt} 
\stackrel{\rho \downarrow 0 }{ \rightarrow}
\left\{ \begin{array}{cc}
{\rm Stop} & {\rm if} {\hspace{0.05in}} 
\sum_{j=2}^{L+1} q_{k,j} \geq \frac{1}{c} \\ 
{\rm Continue} & {\rm if} {\hspace{0.05in}} 
\sum_{j=2}^{L+1} q_{k,j} \leq \frac{1 - h(\rho)}{c} 
\end{array} \nonumber \right. 
\end{eqnarray}
for an appropriately chosen function $h(\rho)$ that satisfies 
$\lim \limits_{\rho \rightarrow 0} h(\rho) = 0$. We start with the 
finite-horizon DP and define ${\bf \Phi}_k$ and ${\bf \Psi}_k$ 
as follows: 
\begin{eqnarray}
{\bf \Phi}_k & \triangleq & \frac{1}{1 + \rho \sum_{j = 2}^{L+1}
q_{k,j}}  - J_k^T(\qb_k), 
{\hspace{0.05in}} 0 \leq k \leq T, \nonumber \\ 
{\bf \Psi}_k & \triangleq & A_k^T(\qb_k) -  \frac{ 1 - \rho} 
{1 + \rho \sum_{j=2}^{L+1} q_{k,j} }  , 
{\hspace{0.05in}} 0 \leq k \leq T-1. \nonumber 
\end{eqnarray}
The main idea behind the proof is to show that ${\bf \Phi}_k$ and 
${\bf \Psi}_k$ are bounded by a function of $\rho$ (that goes to $0$ 
as $\rho \rightarrow 0$), {\em uniformly} for all $k$. Thus, the structure 
of the test in the limit as $\rho \rightarrow 0$ can be obtained. 

Towards this goal, note from Appendix~\ref{app_ccv} that 
${\bf \Phi}_T = {\bf \Psi}_{T-1} = 0$. Also, note that 
$J_{T-1}^T(\qb_{T-1})$ can be written as 
\begin{eqnarray}
J_{T-1}^T(\qb_{T-1}) 
=  \left\{ \begin{array}{cc} 
\frac{1 - \rho + \rho c \sum _{j=2}^{L+1} q_{T-1,j} } 
{1 + \rho \sum_{j=2}^{L+1} q_{T-1,j} } 
& 0 \leq \sum_{j=2}^{L+1} q_{T-1,j} \leq \frac{1}{c} \\ 
\frac{1}{1 + \rho \sum_{j=2}^{L+1}q_{T-1,j} } 
& \sum_{j=2}^{L+1}q_{T-1,j} \geq \frac{1}{c} , 
\end{array}
\right. \nonumber 
\end{eqnarray} 
which can be equivalently written as 
\begin{eqnarray}
{\bf \Phi}_{T-1} = \rho \cdot \frac{1 - c \sum_{j=2}^{L+1} q_{T-1,j} } 
{1 + \rho \sum_{j=2}^{L+1} q_{T-1,j}  } 
\cdot \indic \left( \left\{ \sum_{j=2}^{L+1} q_{T-1, j} 
\leq \frac{1}{c} \right\} \right ). \nonumber
\end{eqnarray} 
Note that $0 \leq {\bf \Phi}_{T-1} \leq \rho$ and we have 
\begin{eqnarray}
0 & \leq & E[  {\bf \Phi}_{T-1} | I_{T-2} ] \triangleq  
{\hspace{0.05in}} - {\bf \Psi}_{T-2} =  \rho g_2(\rho) 
{\hspace{0.07in}} {\rm where} 
\nonumber \\ 
g_2(\rho) 
& \triangleq & 
E \Bigg[ \underbrace{ \frac{1 - c \sum_{j=2}^{L+1} q_{T-1,j} } 
{ 1 + \rho \sum_{j=2}^{L+1} q_{T-1,j} } 
\cdot \indic \left( \left\{ \sum_{j=2}^{L+1} q_{T-1,j} 
\leq \frac{1}{c} \right\}  \right) }_{ X_{\rho} } 
\Bigg| I_{T-2} \Bigg]. \nonumber 
\end{eqnarray} 
Now observe that $X_{\rho}$ can be rewritten as 
\begin{eqnarray}
X_{\rho} 
& = & 
\frac{1 - c \sum_{j=2}^{L+1} q_{T-1,j} } 
{ 1 + \rho \sum_{j=2}^{L+1} q_{T-1,j} } \cdot 
\indic \left( \left\{  p_{T-1,1} \geq \frac{c}{c + \rho} 
\right\} \right). 
\nonumber 
\end{eqnarray}
Furthermore, $X_{\rho} \leq 1$ for all $\rho$ and the set within the 
indicator function (above) converges to the empty set as 
$\rho \downarrow 0$. Thus, a straightforward consequence of the 
bounded convergence theorem for conditional expectation~\cite{durrett} 
is that 
\begin{eqnarray}
\lim_{\rho \downarrow 0} g_2(\rho) & = & 0 \nonumber \\ 
\frac{ {\bf \Psi}_{T-2} }{\rho} & \stackrel{ \rho \downarrow 0 } 
{\rightarrow} & 0, \nonumber
\end{eqnarray}
independent of the choice of $T$.

Plugging the above relation in the expression for $J_{T-2}^T(\qb_{T-2})$, 
we have 
\begin{eqnarray} 
J_{T-2}^T(\qb_{T-2}) 
&=&   \min \Bigg\{  
\frac{1}{1+ \rho \sum_{j=2}^{L+1} q_{T-2,j} }, {\hspace{0.05in}} 
\frac{1 - \rho + \rho c  \sum_{j=2}^{L+1} q_{T-2,j} } 
{  1 + \rho \sum_{j=2}^{L+1} q_{T-2,j} } + 
{\bf \Psi}_{T-2}  
\Bigg\} \nonumber \\ 
& 
= &  \min \Bigg\{  
\frac{1}{1+ \rho \sum_{j=2}^{L+1} q_{T-2,j} }, 
\frac{1 - \rho \left(1 - \frac{ {\bf \Psi}_{T-2}  }{\rho}  \right) 
+ \rho c \sum_{j=2}^{L+1} q_{T-2,j} 
\left( 1 + \frac{ {\bf \Psi}_{T-2} }{c} \right)
} {  1 + \rho \sum_{j=2}^{L+1} q_{T-2,j} } \Bigg\} \nonumber \\ 
& 
= &
\frac{1}{1 + \rho \sum_{j=2}^{L+1} q_{T-2,j} }  
- {\bf \Phi}_{T-2} \nonumber \\ 
{\bf \Phi}_{T-2} & = &  \frac{\rho - {\bf \Psi}_{T-2} 
- \rho \left( c + {\bf \Psi}_{T-2}  \right) \sum_{j=2}^{L+1} 
q_{T-2,j}  } 
{1 + \rho \sum_{j=2}^{L+1} q_{T-2,j} } 
\cdot 
\indic \left( \left\{ \sum_{j=2}^{L+1} q_{T-2,j} 
\leq \frac{1}{c} \cdot 
\frac{ 1 - \frac{ {\bf \Psi}_{T-2} }{\rho}  }  
{ 1 + \frac{{\bf \Psi}_{T-2} }{c}  } 
\right\} \right) 
\nonumber \\
& = & \rho \cdot \left[ \frac{1 - c \sum_{j=2}^{L+1} q_{T-2,j}  } 
{ 1 + \rho \sum_{j=2}^{L+1} q_{T-2,j} } + g_2(\rho) \right]
\cdot \indic \left( \left\{ p_{T-2,1} \geq \frac{c - \rho g_2(\rho)} 
{c+\rho} \right\} \right) \nonumber 
\end{eqnarray} 
with $0 \leq {\bf \Phi}_{T-2} \leq \rho (1 + g_2(\rho))$. As before, 
it is straightforward to check that the set within the indicator 
function converges to the empty set as $\rho \downarrow 0$ and we 
can write ${\bf \Psi}_{T-3}$ as 
\begin{eqnarray}
- {\bf \Psi}_{T-3} & = & E \left[ {\bf \Phi}_{T-2} | I_{T-3} \right] 
= \rho g_3(\rho) \nonumber \\
g_3(\rho) &= & E \Bigg[ \left( \frac{1 - c \sum_{j=2}^{L+1} q_{T-2,j} } 
{1 + \rho \sum_{j=2}^{L+1} q_{T-2,j} } + g_2(\rho) \right)
\cdot \indic \left( \left\{  p_{T-2,1} \geq \frac{c - \rho g_2(\rho)} 
{c+\rho}  \right\} \right) \Bigg| I_{T-3} \Bigg] {\hspace{0.07in}} 
{\rm with} \nonumber \\ 
\lim_{\rho \downarrow 0} g_3(\rho) & = & 0 
{\hspace{0.1in}} {\rm and} {\hspace{0.1in}} 
\frac{ {\bf \Psi}_{T-3} }{\rho} \stackrel{ \rho \downarrow 0 } 
{\rightarrow}  0. \nonumber
\end{eqnarray}

Following the same logic inductively, it can be checked that 
\begin{eqnarray}
\frac{  {\bf \Psi}_{T-k} }{\rho} \stackrel{ \rho \downarrow 0 }{\rightarrow} 0 , 
{\hspace{0.07in}} 1 \leq k \leq T, 
\nonumber 
\end{eqnarray} 
independent of the choice of $T$. That is, we have 
\begin{eqnarray} 
& J_{k}^T(\qb_{k}) 
=   \min \Bigg\{  
\frac{1}{1+ \rho \sum_{j=2}^{L+1} q_{k,j} }, {\hspace{0.05in}} 
\frac{1 - \rho + \rho c \sum_{j=2}^{L+1} q_{k,j} } 
{  1 + \rho \sum_{j=2}^{L+1} q_{k,j} } + 
{\bf \Psi}_{k}  
\Bigg\}. \nonumber 
\end{eqnarray}
Thus, the test structure reduces to stopping when 
\begin{eqnarray}
\sum_{j = 2}^{L+1} q_{k,j } \geq \frac{1}{c} \cdot 
\frac{ 1 - \frac{ {\bf \Psi}_k }{\rho}  } 
{ 1 + \frac{ {\bf \Psi}_k }{c} }, \nonumber
\end{eqnarray}
and using the limiting form for ${\bf \Psi}_k$ as $\rho \rightarrow 0$, 
we have the threshold structure (as stated). The proof is complete 
by going from the finite-horizon DP to the infinite-horizon version as 
in the proof of Theorem~\ref{thm_structure}. Note that while we expect the 
limiting test structure in the finite-horizon setting to be dependent on 
$T$, it is not seen to be the case in this work because $\rho = 0$ is a 
discontinuity point for the DP. 
\endproof

\subsection{Proof of Proposition~\ref{prop_init}} 
\label{app_prop_init}
We first intend to show that a version of~\cite[Lemma 1]{TVAS} holds 
in our case. More precisely, our goal is to show that for any 
$\epsilon \in (0,1)$, we have 
\begin{eqnarray}
\lim \limits_{ \alpha \rightarrow 0 } \sup \limits_{\tau 
{\hspace{0.01in}} \in {\hspace{0.01in}} {\bf \Delta}_{\alpha} } 
P_k \big( \{ k \leq \tau < k + (1 - \epsilon) L_{\alpha} 
\} \big) = 0, \nonumber 
\end{eqnarray}
where $P_k\big(\{ \cdot \}\big)$ denotes the probability measure 
when $\Gamma_1 = k$ and 
\begin{eqnarray}
L_{\alpha} \triangleq \frac{ \log \left( \frac{1}{\rho \alpha} \right) } 
{ L D(f_1, f_0) + |\log(1 - \rho) |}. \nonumber 
\end{eqnarray}
Note that $L_{\alpha} \rightarrow \infty$ as $\alpha \rightarrow 0$. 
Following along the logic of the proof of~\cite[Lemma 1]{TVAS} here, 
it can be seen that 
\begin{eqnarray}
P_k \big( \{ k \leq \tau < k + (1 - \epsilon) L_{\alpha} 
\} \big) & \leq & 
\exp \left( (1 - \epsilon^2) qL_{\alpha} \right)
P_{\infty} \big( \{ k \leq \tau < k + (1 - \epsilon) L_{\alpha}
\}  \big) \nonumber \\ 
& & {\hspace{0.1in}} 
+ P_k \big( \{ \max \limits _{0 \leq n 
< (1 - \epsilon) L_{\alpha} } {\cal Z}_{k+n}^k \geq (1 - \epsilon^2) q 
L_{\alpha} \} \big), 
\label{firstterm}
\end{eqnarray}
where $q \triangleq 
L D(f_1, f_0)$, $P_{\infty}\big(\{ \cdot \}\big)$ denotes the 
probability measure when no change happens, and 
\begin{eqnarray}
{\cal Z}_{k+n}^k = \sum_{\ell = 1}^L 
\sum_{i = \Gamma_{\ell}}^{ k + n} 
\log \left( \frac{f_1(  Z_{i,\ell} ) }{ f_0( Z_{i,\ell} ) } 
\right) \nonumber 
\end{eqnarray}
with $\Gamma_1 = k$. 

For the first term in~(\ref{firstterm}), we have the following. With 
the appropriate definitions of $q$ and $L_{\alpha}$, and the tail 
probability distribution of a geometric random variable, it is again 
easy to check (as in the proof of Lemma 1) that for any $\tau \in 
{\bf \Delta}_{\alpha}$, we have 
\begin{eqnarray}
\exp \left( (1 - \epsilon^2) qL_{\alpha} \right)
P_{\infty} \big( \{ k \leq \tau < k + (1 - \epsilon) L_{\alpha}
\}  \big) \rightarrow 0 {\hspace{0.08in}} {\rm as} 
{\hspace{0.08in}} \alpha \rightarrow 0 \nonumber 
\end{eqnarray}
for any $\epsilon \in (0,1)$ and all $k \geq 1$. For the second 
term in~(\ref{firstterm}), we need a condition analogous 
to~\cite[eqn.\ (3.2)]{TVAS}: 
\begin{eqnarray}
P_k \left( \left\{ \frac{1}{M} \max \limits_{0 \leq n < M } 
{\cal Z}_{k+n}^k \geq (1 + \epsilon) q  \right\} \right) 
\stackrel{M \rightarrow \infty}{ \rightarrow } 0 
{\hspace{0.07in}} {\rm for} {\hspace{0.05in}} {\rm all} 
{\hspace{0.05in}} \epsilon > 0 {\hspace{0.05in}} {\rm and} 
{\hspace{0.05in}} k \geq 1. \nonumber 
\end{eqnarray}
This is trivial since the following is true: 
\begin{eqnarray}
\frac{ {\cal Z}_{k+n}^k }{n } \stackrel{a.s.}{ \rightarrow} 
L D(f_1, f_0) {\hspace{0.08in}} {\rm as} {\hspace{0.08in}} 
n \rightarrow \infty 
\label{cond3pt1}
\end{eqnarray}
for all $k \in [1, \infty)$. 

The above condition follows from the following series of steps. 
First, note that the strong law of large numbers for i.i.d.\ 
random variables implies that 
\begin{eqnarray}
\frac{ {\cal Z}_{k+n}^k }{n } + \frac{1}{n} 
\sum_{\ell = 2}^L \underbrace{ 
\sum_{i = \Gamma_1}^{ \Gamma_{\ell } -1 } 
\log \left( \frac{ f_1(Z_{i,\ell} )  } 
{f_0(Z_{i,\ell})} \right) }_{z_{\ell}} 
& \stackrel{ a.s. } {\rightarrow} & L D(f_1, f_0) 
= q {\hspace{0.08in}} {\rm as} 
{\hspace{0.08in}} n \rightarrow \infty . \nonumber 
\end{eqnarray} 
Then, it can be easily checked that 
\begin{eqnarray}
E \left[ z_{\ell} \right] & = & D(f_1,f_0) 
\sum_{j = 2}^{\ell} \frac{ \left(1 - \rho_{j-1, j} \right)  } 
{\rho_{j-1,j}}. 
\nonumber 
\end{eqnarray}
Since $\min \limits_{\ell} \rho_{\ell-1,\ell} > 0$ from the statement 
of the proposition, we have $E[z_{\ell}] \in (0, \infty)$ for all 
$\ell = 2, \cdots, L$, and hence, the condition in~(\ref{cond3pt1}) holds. 
Applying the condition in~\ref{cond3pt1} with $M = (1 - \epsilon)L_{\alpha}$ 
as $\alpha \rightarrow 0$, we have the equivalent of~\cite[Lemma 1]{TVAS}. 

The proposition follows by application of an equivalent version 
of~\cite[Theorem 1, eqn.\ (3.14)]{TVAS} which follows exactly as 
in~\cite{TVAS}. 
\endproof

\subsection{Completing Proofs of Statements in Sec.~\ref{sectionxxx}} 
\label{app_completionx}

\noindent{\bf \em Proof of Prop.~\ref{prop_reduce}:} We start 
from~(\ref{r_kell_recursion}) and apply the recursion relationship for 
$\{  q_{k-1,\ell}\}$. Noting that $w_m^j w_j^{\ell} = w_m^{\ell}$ for all 
$j$ such that $m \leq j \leq \ell$, we can collect the contributions of 
different terms and write $\sum_{j=1}^{\ell} q_{k-1,j} \hspp w_{j}^{\ell}$ 
as 
\begin{eqnarray}
\sum_{j=1}^{\ell} q_{k-1,j} \hspp w_{j}^{\ell} & = &
\frac{1}{1 - \rho } \cdot \sum_{j = 1}^{\ell} 
q_{k-2, j} \hspp w_{j}^{\ell} \hspp B_{k-1, j,\ell}  
\nonumber 
\end{eqnarray}
where $\{ B_{k-1, j,\ell} \}$ is as defined in the statement of 
the proposition. Thus, we have 
\begin{eqnarray}
\sum_{j=1}^{\ell} q_{k-1,j} \hspp w_{j}^{\ell} & = &
\frac{ (1 - \rho_{\ell-1,\ell}) \prod_{j=1}^{\ell-1}L_{k-1, j } 
}{1 - \rho } \cdot 
\left(\sum_{j = 1}^{\ell} q_{k-2, j} \hspp w_{j}^{\ell} \right) \cdot 
\left\{1 + \zeta_{k-2, \ell} \right\}
\nonumber \\ 
\zeta_{k-2, \ell} & = & 
\frac{1}{(1 - \rho_{\ell-1,\ell}) \prod_{j=1}^{\ell-1} L_{k-1, j } } 
\cdot 
\frac{ \sum_{j=1}^{\ell-1} q_{k-2,j} \hspp w_j^{\ell} \hspp 
C_{k-1, j,\ell} }
{ \sum_{j=1}^{\ell} q_{k-2,j} \hspp w_j^{\ell}  } . 
\nonumber 
\end{eqnarray}
Iterating the above equation, 
we have the conclusion in the statement of the proposition. 

It is useful to reduce Prop.~\ref{prop_reduce} to the case 
of~\cite{venu_qdecentral} when $\rho_{\ell-1,\ell} = 1$ for all 
$\ell = 2, \cdots, L$. For this, note that $\alpha_{k,\ell}$ 
(and hence, $q_{k,\ell}$) are identically zero for all $2 \leq \ell \leq L$. 
Thus, we have 
\begin{eqnarray}
q_{k, L+1} = \alpha_{k,L+1} \cdot \prod_{j=1}^L \prod_{m=1}^k L_{m,j} 
\cdot \prod_{m=0}^{k-2} \left( 1 + \zeta_{m,L+1}  \right). 
\nonumber 
\end{eqnarray}
We then have the following reductions: 
\begin{eqnarray}
\alpha_{k,L+1} & = & \frac{ 1  }{ (1 - \rho)^k } \cdot 
\left(1 + \frac{1}{1 - \rho}  \right). 
\nonumber \\ 
\zeta_{m, L+1} & = & \frac{1}{ \prod_{j=1}^L L_{m+1, j} } \cdot 
\frac{ B_{m+1, 1, L+1} } { 1 + q_{m, L+1} } \nonumber \\ 
B_{m+1, 1, L+1} & = & 1 - \rho {\hspace{0.1in}} {\rm and} 
{\hspace{0.1in}} {\rm hence}, \nonumber \\ 
q_{k,L+1} &= & \frac{ \prod_{j=1}^L L_{k,j} } {1 - \rho} 
\cdot \prod_{m = 0}^{k-1} \left\{  \frac{1}{1 + q_{m-1, L+1}} 
+ \frac{ \prod_{j=1}^L L_{m,j} } {1 - \rho}  \right\} 
\nonumber \\
& = & \frac{ \prod_{j=1}^L L_{k,j} } {1 - \rho} 
\cdot \frac{1}{ \prod_{m=-1}^{k-2} (1 + q_{m, L+1} ) } \cdot 
\prod_{m = 0}^{k-1} \left\{ 1 + \frac{\prod_{j=1}^L L_{m,j} (1 + q_{m-1,L+1}) 
}{1 - \rho}  \right\} \nonumber 
\end{eqnarray}
with the initial condition that $q_{-1,L+1} = 0$ and $L_{0,j} = 1$ for all 
$j$. It is straightforward to establish via induction that the only way in 
which the above recursion can hold is if $q_{k,L+1}$ satisfies 
\begin{eqnarray}
q_{k,L+1} 
& = & 
\frac{ \prod_{j=1}^L L_{k,j} } {1 - \rho} \cdot (1 + q_{k-1,L+1}) 
\nonumber 
\end{eqnarray}
which, as expected, is the same recursion as~(\ref{qk_rec}). 
\endproof 

\noindent {\bf \em Proof of Prop.~\ref{prop_infty_bd}:}
First, note that if we can find $\{ U_k \}$ such that for all $k$ 
\begin{eqnarray}
\log \left( \sum_{\ell =2}^{L+1} \alpha_{k,\ell} \cdot 
C_1 \cdots C_{\ell-1} \cdot J_{\ell} \right) 
\leq U_k, \nonumber  
\end{eqnarray}
then $\nu_{ A  } \geq \nu_{U, A}$ where 
\begin{eqnarray}
\nu_{U,A} \triangleq  \inf_k \Big\{ U_k > A  \Big\}. \nonumber 
\end{eqnarray}
We use Lemma~\ref{lem_bd} to obtain the following bound and the 
associated $\{ U_k \}$: 
\begin{eqnarray}
\sum_{\ell =2}^{L+1} 
\alpha_{k,\ell} \cdot C_1 \cdots C_{\ell-1} \cdot J_{\ell} 
& \leq & 
\sum_{\ell = 2}^{L+1} \frac{ (1 - \rho_{\ell - 1,\ell}) \cdot 
\prod_{j=1}^{\ell-1} L_{k,j} \cdot D_{\ell} } 
{1 - \rho} \cdot 
\prod_{m=1}^{k-1} \frac{ 
\sum_{p = 0}^{\ell-1} 
\big(1 - \rho_{p, p+1}\big) \prod_{j=1}^p L_{m,j} }{1 - \rho} \nonumber \\ 
& \leq & \frac{1}{1 - \rho} \cdot \left( 
\sum_{\ell = 2}^{L+1} D_{\ell} \cdot \prod_{j = 1}^{\ell - 1} L_{k,j} 
\right) \cdot \prod_{m = 1}^{k-1} \frac{
\sum_{p = 0}^{L} (1 - \rho_{p, p+1} ) \prod_{j = 1}^p L_{m,j}} {1 - \rho} 
\nonumber \\ 
& \leq & \frac{D}{1 - \rho} \cdot \left( 
\sum_{p = 1}^{L} \frac{1 - \rho_{p, p+1}}
{1 - \rho} \cdot \prod_{j = 1}^p L_{k,j} \right) 
\cdot \prod_{m = 1}^{k-1} \frac{
\sum_{p = 0}^{L} (1 - \rho_{p, p+1} ) \prod_{j = 1}^p L_{m,j}} {1 - \rho} 
\nonumber \\ 
& \leq & \frac{D}{1 - \rho} \cdot 
\prod_{m = 1}^{k} \frac{
\sum_{p = 0}^{L} (1 - \rho_{p, p+1} ) \prod_{j = 1}^p L_{m,j}} {1 - \rho} 
\nonumber 
\end{eqnarray}
where $D_{\ell} = \prod_{j = 1}^{\ell-2} \rho_{j, j+1} \cdot \left( 
\sum_{j = 0}^{\ell - 1} \frac{ 1 - \rho_{j, j+1} }{1 - \rho} \right)$, 
$D = 1 + \max \limits_{\ell = 1, \cdots, L} 
\frac{\ell}{1 - \rho _{\ell, \ell +1} }$. 
With the above bound, we have 
\begin{eqnarray}
\nu_{ A } \geq \inf_k \left\{ 
\sum_{m = 1}^{k} 
\log \left( 
\frac{ \sum _{p = 0}^{L} 
\big(1 - \rho_{p, p+1}\big) \prod \limits_{j=1}^p L_{m,j} } 
{1 - \rho} \right) > A + \log \left( \frac{1 - \rho}{D} 
\right) \right\}. \nonumber 
\end{eqnarray}
The conclusion follows by using Lemma~\ref{lem_blackwell} and noting 
that $E\left[ \log \left( 
\frac{ \sum_{p = 0}^{L} 
\big(1 - \rho_{p, p+1}\big) \prod_{j=1}^p L_{m,j} } 
{1 - \rho} \right) \right] \in (0,\infty)$. 
\endproof

\noindent {\bf \em Proof of Prop.~\ref{thm_mean_taua}:}
This proof is a formal write-up of the heuristic presented before the 
statement of Prop.~\ref{thm_mean_taua}. Following the definition of $\eta_j$ 
and the fact that $0 \leq \eta_j \leq 1$, we have 
\begin{eqnarray}
\eta_j x_j \leq \prod_{m = \ell^{\star} }^j x_m, 
{\hspace{0.05in}} j \geq \ell^{\star}. 
\nonumber 
\end{eqnarray} 
Suppose there exists an $\ell^{\star} \leq L$ as defined in~(\ref{ellstar}), 
invoking Lemma~\ref{lem_positive} with the fact that 
$\Delta_{\ell^{\star}, j} \leq 0$ for all $j \geq \ell^{\star}$, 
we have 
\begin{eqnarray} 
\frac{1}{k} \sum_{\ell = \ell^{\star}}^L 
\log \left( 1 + \eta_{\ell} x_{\ell} \right) \stackrel{k \rightarrow \infty} 
{\rightarrow} 0 
{\hspace{0.08in}} {\rm a.s.} {\hspace{0.05in}} {\rm and} 
{\hspace{0.05in}} {\rm in} {\hspace{0.05in}} {\rm mean}. 
\nonumber 
\end{eqnarray}
Thus, we have 
\begin{eqnarray} 
\frac{1}{k} \sum_{\ell = 2}^L \log \left(1 +\eta_{\ell } x_{\ell} \right) 
- 
\frac{1}{k}  \sum_{\ell = 2}^{\ell ^{\star} -1 } \log \left(1 + 
\eta_{\ell} x_{\ell}  \right) 
\stackrel{k \rightarrow \infty}{\rightarrow } 
0 {\hspace{0.08in}} {\rm a.s.} {\hspace{0.05in}} {\rm and} 
{\hspace{0.05in}} {\rm in} {\hspace{0.05in}} {\rm mean}. \nonumber
\end{eqnarray}

The main contribution to~(\ref{main_term}) is now established via 
induction. Since $\eta_2 = 1$, we can expand the sum as (modulo the 
a.s.\ and in mean convergence parts): 
\begin{eqnarray} 
\frac{1}{k} \sum_{\ell = 2}^{\ell^{\star} - 1} 
\log \left( 1 + \eta_{\ell} x_{\ell}  \right) - 
\frac{1}{k} \log \left( 1 + \sum_{\ell = 2}^{ \ell^{\star} -1} 
\prod_{m = 2}^{\ell} 
x_m \right) \stackrel{ k \rightarrow \infty }{ \rightarrow } 0 
. \nonumber
\end{eqnarray}
If $\ell^{\star} = 2$, it is clear that the proposition is true. If 
$3 \leq \ell^{\star} \leq L+1$, since $2 < \ell^{\star}$, by the 
definition of $\ell^{\star}$, there exists (a smallest choice) $j_2 \geq 2$ 
such that 
\begin{eqnarray} 
\prod_{m = 2}^{j_2} x_m & \stackrel{k \rightarrow \infty}{ \rightarrow } 
& \infty  {\hspace{0.05in}} {\rm with} \nonumber \\ 
\prod_{m = 2}^p x_m & \stackrel{ k \rightarrow \infty}{\rightarrow} & 0 
{\hspace{0.05in}} {\rm or} {\hspace{0.05in}} {\cal O}(1) 
{\hspace{0.05in}} {\rm for} {\hspace{0.05in}} {\rm all} 
{\hspace{0.05in}} 2 \leq p \leq j_2-1 \nonumber 
\end{eqnarray}
provided the set $[2, \cdots, j_2-1]$ is not empty. There are two possibilities: 
$j_2 = \ell^{\star} - 1$ or $j_2 \leq \ell^{\star}-2$. (Note that $j_2 \geq  
\ell^{\star}$ results in a contradiction since it will imply 
$\prod_{m = \ell^{\star} }^{j_2} x_m \rightarrow \infty$, but we know 
this is not true from the definition of $\ell^{\star}$). 
In the first case, we are done upon invoking Lemma~\ref{lem_positive}. 
In the second case, 
iterating by replacing $2$ with $j_2 + 1$ (as many times as necessary) 
and finally invoking Lemma~\ref{lem_positive} and noting the main contribution 
of the sum in~(\ref{main_term}), we arrive at the conclusion of the 
proposition. 
\endproof

\bibliographystyle{IEEEbib}
\bibliography{newrefsx}
\end{document}